\documentclass[aps,showpacs]{revtex4}
\usepackage{amssymb}
\usepackage{graphicx}
\usepackage{euscript}
\usepackage{pstricks}

\begin{document}
\title{\bf Light propagation with non-minimal couplings
in a two-component cosmic dark fluid with an Archimedean-type force
and unlighted cosmological epochs}

\author{Alexander B. Balakin\footnote{Email:
Alexander.Balakin@ksu.ru}}
\affiliation{Kazan Federal University, Institute of Physics, \\ 
Kremlevskaya str.,
18, 420008, Kazan,  Russia,}
\author{Vladimir V. Bochkarev\footnote{Email: Vladimir.Bochkarev@ksu.ru}}
\affiliation{Kazan {} Federal University, Institute of Physics, \\ 
Kremlevskaya str.,
18, 420008, Kazan,  Russia,}
\author{Jos\'e P. S. Lemos\footnote{Email: joselemos@ist.utl.pt}}
\affiliation{Centro Multidisciplinar de Astrof\'{\i}sica -
CENTRA, Departamento de F\'{\i}sica, Instituto Superior T\'ecnico - IST,
Universidade T\'ecnica de Lisboa - UTL,
Av. Rovisco Pais 1, 1049-001 Lisboa, Portugal.}

\begin{abstract}

During the evolution of the universe there are at least two epochs
during which electromagnetic waves cannot scan the universe's internal
structure neither bring information to outside observers. The first
epoch is when photons are in local thermodynamic equilibrium with
other particles, and the second is when photon scattering by charged
particles is strong. One can call these two periods of cosmological
time as standard unlighted epochs.  After the last scattering surface,
photons become relic photons and turn into a source of information
about the universe.
Unlighted cosmic epochs can also appear when one considers non-minimal
theories, i.e., theories in which the electromagnetic field is coupled
in an intricate way with the cosmological gravitational field.  By
considering a cosmological model where the dark sector, i.e., the
dark energy and dark matter, self-interacts via an
Archimedean-type force, and taking into account a non-minimal coupling
theory for the electromagnetic field, we discuss the appearance of
unlighted epochs. 
In the framework of our non-minimal theory, a three-parameter
non-minimal Einstein-Maxwell model, the curvature coupling can be
formulated in terms of an effective refraction index $n(t)$. Then,
taking advantage of a well-known classical analogy, namely, in a
medium with $n^2<0$ electromagnetic waves do not propagate and their
group velocity, i.e., energy transfer velocity, has zero value at the
boundary of the corresponding zone, one can search for the unlighted
epochs arising in the interacting dark fluid
cosmological model.  We study here, both
analytically and numerically, cosmological models admitting unlighted
epochs.

\end{abstract}
\maketitle

\newpage

\section{Introduction}

\subsection{The cosmology, two-component
dark fluid and Archimedean-type interaction}

In order to examine important features happening within the universe,
such as light propagation, one has to set up from the start
a model for the
dynamics of the universe as a whole, smoothing out all irregularities.
General relativity tells that both geometry and matter are important
in the dynamics of the universe and, moreover, 
through Einstein's equations one
finds that geometry guides the matter and matter changes the geometry.

In relation to geometry, we now have a good idea of the spacetime
geometry of the universe. It is governed by a cosmological time $t$,
which is the proper time of the fundamental particles, usually
considered as galaxies, of the substratum. Slices of this time yield,
by the use of the cosmological principle, a homogeneous and isotropic
spatial geometry, so that the metric is 
the Friedmann-Lema\^itre-Robertson-Walker (FLRW) metric.
From observations the universe is expanding, which is
then taken into account by a single function, the cosmological scale
factor $a(t)$, which once known, also yields the rate of expansion,
i.e., the Hubble
function $H(t)\equiv \frac{\dot{a}}{a}$, and the rate
of acceleration, i.e., the acceleration parameter
${-}q(t)\equiv \frac{\ddot{a}}{a H^2}$.
Moreover, the spatial slices are flat, or almost flat,
simplifying the problem even further (see, e.g.,
\cite{bookoncosmology}).

In relation to matter, we also have now a good idea of the matter
content of the universe. There are three main components, namely, dark
energy which accounts for 70\% of the matter, dark matter which
accounts for around 25\%, and the rest, which includes baryonic
matter, radiation,  
and other forms such as black holes, 
which accounts for 5\%. Dark energy
and dark mater are essential building blocks in any cosmological
model.  Their precise nature, both of the dark energy and dark matter,
is unknown, but this does not preclude a good understanding of the
gross features of the dynamics of the universe.  Dark energy
\cite{DE1,DE2,DE3} is essential in the explanation of the observed
accelerated expansion of the universe \cite{A1,A2}. Dark matter
provides an excellent explanation for the flat velocity curves seen in
the outskirts of a galaxy and for the dispersion velocities of
galaxies in clusters of galaxies \cite{DM1,DM2}.  Given there are
these two important matter components that essentially drive the
dynamics, it is worth contemplate them as
a single whole. There are various manners in which this could be
performed. One intriguing possibility is considering both components as
manifestations of a single dark fluid \cite{DF1,DF2,DF3,DF4,DF5}.
Other possibility is to postulate an interaction between them so that
the two components although really distinct have to be treated in a
broader unified scheme, see e.g., \cite{zpc,vmm}.

In \cite{Arch1,Arch2} this latter possibility of a unified interaction
scheme between dark energy and dark matter has been followed.  The
background for the interaction itself was given by relativistic
hydrodynamics theory for the dark energy component and a relativistic
kinetics framework for the dark matter component.  Baryonic matter is
negligible in the context of cosmological dynamics and so it has been
left out in the scheme.  The interaction itself between both
components has been provided by an Archimedean-type four-force, a direct
generalization of the Archimedean buoyancy law, in which the
three-force is proportional to the gradient of the pressure.  The dark
energy pressure is of the same order as the dark energy energy density
and for this reason, the influence of the Archimedean-type force on the
dark matter component can be important.  Now, the dark energy pressure
can be negative, in which case the Archimedean-type force can be negative,
instead of positive as in the usual buoyancy force case.  In the case
the pressure changes sign as the universe evolves, there appear
several different stages in its acceleration, in which the
acceleration itself can change sign. It was further shown in
\cite{Arch1,Arch2} that the Archimedean-type force is able to distribute
between the two fluid components the corresponding energy content
of the universe, which in turn guides the whole evolution of the
universe, in its one or several stages of acceleration.
Multistage universes have also appeared in \cite{pen}.

\subsection{Non-minimal coupling}

The way in which the electromagnetic field couples to gravity is an
open question. Usually it is assumed a minimal coupling where simple
flat spacetime derivatives involving
the field are replaced by covariant
derivatives but this might not be so. For instance, one can replace
the flat spacetime 
derivatives by covariant derivatives along with terms
involving the curvature tensor and its contractions, yielding a
non-minimal coupled theory. In a gravitational 
strong regime, like in the vicinity
of black holes or in a cosmological context, these additional terms,
if present, can be felt and are important
(see, e.g., original papers and reviews
\cite{1,Horn,Drummond,Novello2,Acci1,Souza,Go,Acci2,Turner,MHS,Lafrance,B1,Acci3,Mohanty,FaraR,hehl,Kost1,Kost2,Prasa3,Tess,Solanki,6}).

A simple and most fruitful non-minimal theory, a non-minimal
Einstein-Maxwell theory, possesses an action with a Lagrangian which
includes a linear combination of three cross-invariants, namely, $q_1
R F_{mn}F^{mn}$, $q_2 R^{ik}F_{im}F_{kn}g^{mn}$ and $q_3
R^{ikmn}F_{ik}F_{mn}$, where $q_1$, $q_2$ and $q_3$ are free
parameters. The scalars appearing in the Lagrangian are linear in the
Riemann tensor $R^i_{\ kmn}$, Ricci tensor $R_{kn}$, and Ricci scalar
$R$, and are quadratic in the Maxwell tensor $F_{ik}$.  An interesting
fact about this non-minimal theory is that the corresponding
non-minimal electrodynamic equations have the same form as the
equations for the electrodynamics of anisotropic inhomogeneous
continuous media (see, e.g., \cite{EM,HehlObukhov,bl05}).  This
circumstance makes it possible to consider non-minimal analogs of
well-known phenomena in classical electrodynamics of continuous media,
such as birefringence induced by curvature \cite{B1}, anomalous
behavior of electric and magnetic fields non-minimally coupled to the
gravitational radiation \cite{BL1}, curvature induced Cherenkov
effects \cite{BKL}, and optical activity in vacuum \cite{BL2}.  Of
course, the variety of non-minimal effects depends on the spacetime of
the model and its symmetries.

Since electromagnetic waves, from radio
waves, to light, to gamma rays, are a very important source of
information about the universe, it is worth to study the changes that
might operate in a universe with a non-minimal coupling between the
electromagnetic and gravitational fields.

\subsection{Non-minimal electromagnetic wave propagation in
an expanding universe with a cosmic
dark fluid with Archimedean-type interaction and unlighted epochs}

\subsubsection{Prologue}

In a generic gravitational background field, electromagnetic waves
non-minimally coupled propagate with a velocity which differs from the
velocity of light in vacuum. These waves, influenced by tidal
interactions induced by curvature, do not travel along null geodesics
of the background spacetime. Thus, non-minimal coupling may produce
significant changes in the propagation of electromagnetic waves. It is
therefore worth to study this coupling in a cosmological context.

In cosmology, in particular, when one deals with a spatially
homogeneous isotropic expansion for the universe, one is faced with
electromagnetic effects, possibly non-minimal, of two types: first,
the phase and group wave velocities depend on the cosmological time,
and second, the amplitude and energy density of the electromagnetic
waves decrease with time. Different aspects of these two phenomena
have already been studied within a non-minimal setting. For instance,
in \cite{Tess} the deviation of the photon velocity from the velocity
of light in vacuum was estimated in the framework of Drummond and
Hathrell's approach \cite{Drummond}.

\subsubsection{Motivation}

We are interested here in developing the study of the propagation of
electromagnetic waves non-minimally coupled to the gravitational field
in an isotropically expanding universe. Since cosmological models with
dark fluid interaction are of great interest, such as those studied in
\cite{Arch1,Arch2}, there are now new motives to study anew the
propagation of these electromagnetic waves non-minimally coupled to
the gravitational field in an isotropically expanding universe
governed by a dark fluid. We mention three such motives.

The first motive is related with the synergy between non-minimal
electromagnetic wave propagation and a cosmology in which a dark fluid
(i.e., a fluid with two components, namely, dark energy plus dark
matter) plays a major role \cite{Arch1,Arch2}.  In our non-minimal
theory there are three coupling constants $q_1$, $q_2$ and
$q_3$. These non-minimal constants can be determined directly from a
more fundamental theory, as was done in the case of quantum
electrodynamics vacuum polarization effects in a curved background by
Drummond and Hathrell \cite{Drummond}, or they can be considered as
phenomenological inputs.  Of course, the results for $q_1$, $q_2$ and
$q_3$, in \cite{Drummond} should be thought as providing a portion of
the whole values to the three constants, since other effects can
provide curvature induced interactions between the fields, and all the
effects should be summed together. Our idea here is that a dark fluid
can also support non-minimal interactions. Thus, we consider the
parameters $q_1$, $q_2$ and $q_3$ to be phenomenologically introduced
and do trust that a mechanism of non-minimal coupling between
electromagnetism and gravity via a dark fluid can be clarified in the
future. In other words, if a dark fluid, as a medium with unusual
properties, can act as a mediator of non-minimal coupling between
photons and gravitons, we expect that the coupling parameters $q_1$,
$q_2$ and $q_3$ will be estimated theoretically and established from
observations.

The second motive is related to the existence of unlighted epochs in
our models.  The non-minimal coupling of the electromagnetic waves
with the gravitational field can be codified in terms of an effective
time-dependent refraction index $n(t)$, where $t$ denotes cosmological
time. From this, we will show that for specific values of the
non-minimal parameters $q_1$, $q_2$ and $q_3$, $n^2(t)$ can be
negative during some finite time interval, and can either vanish or
take infinite values at some transition points $t_{(s)}$, i.e., points
where the sign of $n^2(t)$ changes, the subscript $s$, labeling the
several possible transition points $s=1,2,...$ and denoting sign
change. This means that the universe can pass through epochs when
electromagnetic waves cannot propagate as their phase velocity is a
pure imaginary quantity during this period of time. Taking into
account that in order to read the history of the universe one should
reconstruct the sequence of the events by using the whole spectrum of
electromagnetic radiation, we coin such epochs as unlighted epochs,
since portraits of the universe of those epochs cannot be available.
Moreover, the equations for trapped surfaces in the universe have here
the form $t=t_{(s)}$, and they are time-like in contrast to the
standard space-like ones that appear in black hole formation. The use
of the term unlighted epoch is then well fit, since it distinguishes
clearly the two situations, namely unlighted epochs in the universe
versus trapped surfaces in black holes.

The third motive is determined by the necessity to divide properly the
history of the universe into epochs according to various physical
scenaria. Since the transition points in the universe history are
fixed by the critical temperatures $T_{(c)}$ of some basic physical
processes, one should link the temperature evolution $T(t)$ and the
cosmological time $t$ by using the cosmological scale factor
$a(t)$. When the cosmological time scale is defined upon using relic
cosmic microwave background photons traveling along null geodesic
lines, the corresponding law is very simple, $T(t)a(t)=T(t_0)a(t_0)$,
where $t_0$ is some reference time, like now, say. On the other hand,
when we take into account a non-minimal coupling between photons and
gravitons, the corresponding link between the temperature $T(t)$ and
the scale factor $a(t)$ is much more sophisticated. Indeed, it is
determined not only by $a(t)$, but also by its first and second
derivatives, $\dot{a}(t)$ and $\ddot{a}(t)$, respectively.  In other
words, the temperature $T(t)$ is tied to the scale factor $a(t)$, the
Hubble function $H(t)\equiv \frac{\dot{a}}{a}$ and the acceleration
parameter ${-}q(t)\equiv \frac{\ddot{a}}{a H^2}$. This fact, of
course, should induce novel and interesting features.

\subsubsection{This work}

In
\cite{Arch1,Arch2} a
cosmological model based on a dark fluid made of two components,
the dark energy, considered as the fluid substratum,
and the
dark matter, described in the framework of kinetic theory,
interacting via an Archimedean-type force
was considered. The
evolution models for the universe were
classified in terms of the admissible
transition points, i.e., the points in which an accelerated expansion
of the universe is changed to a decelerated expansion and vice versa.
This classification is based on the analysis of the function
acceleration parameter
${-}q(t)\equiv \frac{\ddot{a}}{a H^2}$.
Now, we find here that the
function $n^2(t)$, describing the behavior of
the effective refraction index, also includes the functions $H(t)$ and
$-q(t)$.  Thus, in order to analyze non-minimal light propagation,
it is useful to use $n^2(t)$ instead of $-q(t)$.
In this work we study,
analytically and numerically, the propagation
of electromagnetic waves non-minimally coupled,
along with their phase and group velocities,
in six cosmological models based on a dark fluid
which self interacts through an Archimedean-type
force \cite{Arch1,Arch2}.
This work is thus an interesting sequel of \cite{Arch1,Arch2}.

\subsubsection{The organization of the paper}
The paper is organized as follows.  In Section \ref{cosmodel} we
introduce the cosmological model.  In subsection \ref{cosmcont} we
give the cosmological model itself and recall the basic formulas of
the Archimedean-type model, which are necessary for further numerical
analysis.  In subsection \ref{set} we give the basic set up for
non-minimal interaction and light propagation.  In Section
\ref{optmet} using the master equations of non-minimal electrodynamics
we reconstruct the effective dielectric and magnetic permeabilities,
the effective refraction index $n(t)$, the effective (optical) metric
in terms of the scale factor $a(t)$, the Hubble function $H(t)$,
acceleration parameter $-q(t)$ and the three non-minimal coupling
phenomenological constants $q_1$, $q_2$ and $q_3$. We also discuss the
refraction index $n(t)$, focus on the derivation of the expression for 
the group velocity of the waves (i.e., the velocity of energy
transfer) using the analogy with electrodynamics of continua, and
define unlighted epochs.  In section \ref{analyticalstudy} we make an
analytical study of the unlighted epochs. In subsection \ref{choiceag}
using the Kohlrausch stretched exponential functions we give the
cosmological models fit for the study, in subsection \ref{choicesub}
we provide the choice for the non-minimal models, and in \ref{2m} we
give some analytical cosmological examples describing unlighted
epochs.  In Section \ref{num} we consider the results of numerical
analysis for the refraction index $n(t)$, phase and group velocities and
effective lengths of the photon trips for six basic Archimedean-type
submodels.  In subsection \ref{choice} we discuss the cosmology, in
subsection \ref{choicesub2} we provide the choice for the non-minimal
models, and in subsection \ref{6m} the models are analyzed in detail,
including the submodels of perpetually accelerated universe, periodic
and quasiperiodic universes, and various submodels with one, two and
three transition points.  We also study the photon pathlength, the
true duration of epochs and the universe lifetime for these models.
In Section \ref{conc} we draw conclusions.  In the Appendix we analyze
a number of one-parameter examples for which the coupling constants
are linked by two relations motivated physically and geometrically.

\section{The cosmological model}
\label{cosmodel}

\subsection{Cosmological context}
\label{cosmcont}

\subsubsection{The Cosmology}
\label{thecosmology1}

We start from the action functional
\begin{equation} \label{actionstart0}
S = \int d^4x \sqrt{-g} \left(\frac{R}{2\kappa} + L^{\rm
matter} \right)\,,
\label{lagrange0}
\end{equation}
where $R$ is the Ricci scalar, $g$ is the determinant of the
four-dimensional spacetime metric
$g_{ik}$, $L^{\rm
matter}$ is the matter Lagrangian
which we assume includes a cosmological constant
$\Lambda$, and
$\kappa {=} \frac{8\pi G}{c^4}$ with $G$
being Newton's constant and $c$ the velocity of the light.
Using the standard variation procedure with respect to the
metric $g^{ik}$
one can obtain Einstein's equations. Latin
indices run from 0 to 3.
In the above
we assume
that $L^{\rm
matter}$, and so the corresponding energy-momentum tensor,
$T^{\rm matter}_{ik}$, comes from
both the dark energy and dark
matter, which together make up for
most of the total energy of the
universe.
Thus, the master equations for
the gravity field are assumed to be the
usual Einstein's equations,
\begin{equation}
R_{ik} - \frac{1}{2}Rg_{ik} =  \kappa
T^{\rm matter}_{ik} \,, \label{081}
\end{equation}
where $R_{ik}$ is the Ricci tensor,
given
by the contraction of the Riemann curvature tensor $R^l_{imk}$.

We use the spatially homogeneous
flat metric of Friedmann-Lema\^itre-Robertson-Walker (FLRW) type
given by
\begin{equation}
ds^2 = dt^2 -  a^2(t) \left(dx^2 + dy^2 +dz^2 \right) \,,
\label{FLRW}
\end{equation}
where we chose units with $c{=}1$. The energy momentum tensor is
described by the functions $\rho(t)$ and $\Pi(t)$, and $E(t)$ and
$P(t)$.  The functions $\rho(t)$ and $\Pi(t)$ describe
the energy density and pressure of the dark energy, respectively. 
The functions
$E(t)$ and $P(t)$ are the energy density and pressure
of the dark matter, respectively. The cosmological constant $\Lambda$ is
incorporated into the dark energy state functions, i.e., the term
$\frac{\Lambda}{8\pi G}$ is included into $\rho$ and $\Pi$
\cite{Arch1}.
With these assumptions
the equations coming out of 
Eq.~(\ref{081}) for the gravitational field read
\begin{equation}
\left(\frac{\dot{a}}{a}\right)^2
= \frac{8\pi G}{3}(\rho + E) \,, \label{EinREDU1a}
\end{equation}
\begin{equation}
{\left(\frac{\dot{a}}{a}\right)}^{\hskip-0.1cm\dot{}}
  + \left(\frac{\dot{a}}{a}\right)^2 =
- \frac{4\pi G}{3}[ (\rho + E) + 3 (\Pi + P)]
\,,  \label{EinREDU10}
\end{equation}
with the dot denoting a derivative with respect to time.

\subsubsection{Two-component dark fluid:
Archimedean-type interaction between dark energy and dark
matter}
\label{arch}

In order to describe the evolution of several quantities, in particular
of the refraction index $n(t)$, as this is the quantity
we are interested,
we need of the Hubble function $H(t)$,  defined as
\begin{equation}
H(t) \equiv
\frac{\dot{a}}{a}\,,
\label{R2a}
\end{equation}
and the
acceleration parameter ${-}q(t)$,
\begin{equation}
-q(t) \equiv \frac{\ddot{a}}{a H^2}  \,.
\label{R2b}
\end{equation}
Sometimes it is useful to swap ${-}q(t)$ for $\dot{H}$,
given by,
\begin{equation}
\dot{H}(t)= - H^2(t)[1+q(t)]  \,.
\label{doth}
\end{equation}
Now, we obtain
these functions using the Archimedean-type interaction
between the dark energy and dark matter model \cite{Arch1,Arch2}.
The
function $H(t)$ can be found from Eq.~(\ref{R2a}) and
Einstein's equation (\ref{EinREDU1a}), yielding
\begin{equation}
H^2(t) = \frac{8\pi G}{3}(\rho + E) \,.
\label{EinREDU1}
\end{equation}
The acceleration parameter $-q(t)$ can be found from Eq.~(\ref{R2b}) and
Einstein's equation (\ref{EinREDU10}), yielding
\begin{equation}
{-}q(t)  {=} {-}
\frac{1}{2}\left[1{+}3 \left(\frac{\Pi{+}P}{\rho{+}E}\right)
\right] \,. \label{q}
\end{equation}
As for the function $\dot{H}(t)$, one
uses Eq.~(\ref{doth}) and Eqs.~(\ref{EinREDU1})-(\ref{q})
to find
\begin{equation}
\dot{H}(t)= 
- 4\pi G \left(\rho+E+\Pi + P \right)\,. \label{q7}
\end{equation}
Thus, we need to find the following state functions: the energy
density of the dark energy $\rho(t)$ and its pressure $\Pi(t)$, the
energy density of the dark matter $E(t)$ and its pressure $P(t)$. The
search scheme for these quantities is the following.  First of all
introducing a new convenient variable
\begin{equation}
x \equiv \frac{a(t)}{a(t_0)} \,,
\label{newx}
\end{equation}
and using the auxiliary formulas
\begin{equation}
\frac{d}{dt} = xH(x)\frac{d}{dx} , \qquad t{-}t_0 =
\int_0^{\frac{a(t)}{a(t_0)}}
\frac{dx}{x H(x)} \,, \label{x}
\end{equation}
we find, through some manipulation of Einstein's equations
a key equation for
$\Pi(x)$
\begin{equation}
\xi x^2 \Pi^{\prime \prime}(x) {+} x \Pi^{\prime}(x) \left(4 \xi {+}
\sigma \right) {+} 3 (1{+}\sigma)\Pi {+}
3 \rho_0 {=} {\cal J}(x) \,,
\label{key1}
\end{equation}
a second order differential equation, linear in
the first and second derivatives, $\Pi^{\prime} \equiv \frac{d}{dx}\Pi$
and $\Pi^{\prime \prime}$, and nonlinear in the function $\Pi(x)$.
The parameters $\xi$, $\sigma$ and $\rho_0$ are the coupling constants
involved into the assumed linear inhomogeneous equation of state,
namely,
\begin{equation}
\rho(x) = \rho_0 + \sigma \Pi(x) + \xi x \frac{d}{dx} \Pi(x) \,.
\label{simplest0}
\end{equation}
This equation links
the pressure to the energy density of the dark energy (see
\cite{Arch1} for details).  The source-term ${\cal J}(x) ={\cal
J}(x,\Pi{-}\Pi(1),\Pi^{\prime})$ in the right-hand side of
(\ref{key1}) is given by the integral
\begin{equation}
{\cal J}(x) {=} {-} \sum_{({\rm a})} E_{({\rm a})} \frac{\left[x^2
F_{({\rm a})}(x)\right]^{\prime}}{2 x^4} \int_0^{\infty}\frac{y^4 dy \
e^{{-}\lambda_{({\rm a})} \sqrt{1{+}y^2}}}{\sqrt{1{+}y^2 F_{({\rm
a})}(x)}}
\,, \label{key2}
\end{equation}
where we used the definitions
\begin{equation}
F_{({\rm a})}(x) = \frac{1}{x^2} \exp{\{2{\cal V}_{({\rm a})}
[\Pi(1){-}\Pi(x)]\}} \,, \label{FF}
\end{equation}
\begin{equation}
E_{({\rm a})} \equiv \frac{N_{({\rm a})} m_{({\rm a})}
\lambda_{({\rm a})}}{K_2(\lambda_{({\rm a})})}\,,  \quad
\lambda_{({\rm a})} \equiv \frac{m_{({\rm a})}}{k_{\rm
B}\,T_{({\rm a})}} \,, \label{E}
\end{equation}
\begin{equation}
K_{\nu}(\lambda_{({\rm a})}) \equiv \int_0^{\infty} dz \cosh{\nu
z}\; \exp{[-\lambda_{({\rm a})} \cosh z]}   \,. \label{McD}
\end{equation}
Here ${\cal V}_{({\rm a})}$ are the constants of the Archimedean-type
coupling (see \cite{Arch1} for details), $K_\nu(\lambda_{({\rm a})})$ are
the modified Bessel functions of second kind of order
$\nu$ (we are interested in the $\nu=2$ case),
$\Pi(1)\equiv \Pi(t_0)$ is the initial value
of the dark energy pressure,
$k_{\rm B}$ is the Boltzmann
constant, and $T_{({\rm a})}$,
$m_{({\rm a})}$, and $N_{({\rm a})}$, are
the partial temperature, the mass,
and the number of particles
per unit volume of the sort $({\rm a})$, respectively.
When the quantity $\Pi(x)$ is found,
other state functions can be calculated using the integrals
\begin{equation}
E(x) = \sum_{({\rm a})}\frac{E_{({\rm a})}}{x^3} \int_0^{\infty}
y^2 dy \sqrt{1{+}y^2 F_{({\rm a})}(x)} \ e^{{-}\lambda_{({\rm a})}
\sqrt{1{+}y^2}}\,, \label{e(x)}
\end{equation}
\begin{equation}
P(x) = \sum_{({\rm a})}\frac{E_{({\rm a})}}{3x^3} \int_0^{\infty}
\frac{F_{({\rm a})}(x) y^4 dy}{\sqrt{1{+}y^2 F_{({\rm a})}(x)}} \
e^{{-}\lambda_{({\rm a})} \sqrt{1{+}y^2}}\,.  \label{p(x)}
\end{equation}
Thus we have formulated our integration scheme.
From here, we want to work out novel features that appear in
such a cosmological
context.

\subsection{Non-minimal coupling: set up}
\label{set}

Now we want to study light propagation
envisaged as a perturbation in the
background cosmological manifold.
This means that the electromagnetic field
propagates in a given geometry, without modifying
the geometry itself. The electromagnetic field
is a test field. This is well justified since
it is known that the energy density of radiation is
thoroughly negligible in relation to the
dark energy, dark matter and baryonic matter.
We also assume that the field is non-minimally coupled
and write the electromagnetic action functional as,
\begin{equation} \label{actionstart}
S^{\rm electromag} = \frac{1}{4} \int d^4x \sqrt{-g}\,
\left( F_{mn}F^{mn} +  {\cal R}^{ikmn}
F_{ik}F_{mn}\right)\,. \label{lagrange1}
\end{equation}
Using the standard variation procedure with respect to the
electromagnetic potential four-vector $A_i$
one can obtain the non-minimally extended Maxwell
equations  (see, e.g.,
\cite{bl05}, see also \cite{hehl}).  Here
$F_{ik}=\partial_i A_k {-}\partial_kA_i$ is the Maxwell tensor.
Then, the non-minimal Maxwell equations have the following form
\begin{equation}
\nabla_k H^{ik} =0 \,, \quad \nabla_k F^{*ik} = 0 \,, \label{max1}
\end{equation}
where the excitation tensor $H^{ik}$ is linked with the Maxwell tensor
$F_{mn}$ by a linear constitutive equation
\begin{equation}
H^{ik} = C^{ikmn} F_{mn} \,, \label{max21}
\end{equation}
with the linear response tensor $C^{ikmn}$ given by
\begin{equation}
C^{ikmn}= \frac{1}{2}(g^{im}g^{kn} {-} g^{in}g^{km}) + {\cal
R}^{ikmn}\,. \label{max213}
\end{equation}
The dual tensor $F^{*ik} \equiv \frac{1}{2}\epsilon^{ikmn}F_{mn}$ is
defined in a standard way through the Levi-Civita tensor
$\epsilon^{ikmn}$. The tensor ${\cal R}^{ikmn}$ is the non-minimal
susceptibility tensor decomposed as
\begin{equation}
{\cal R}^{ikmn} \equiv \frac{q_1 R}{2}(g^{im}g^{kn} {-} g^{in}g^{km})
{+} \frac{q_2}{2} (R^{im}g^{kn} {-} R^{in}g^{km} { +} R^{kn}g^{im} {-}
R^{km}g^{in}) {+} q_3 R^{ikmn} \,, \label{susceptibility2}
\end{equation}
where $q_1$, $q_2$, and $q_3$ are the phenomenological parameters of
the non-minimal coupling, and
$R^{ikmn}$, $R^{im}$, and $R$, are the Riemann tensor, the Ricci tensor,
and the Ricci
scalar of the background cosmological
manifold, respectively. It is appropriate here to recall that
the contracted susceptibility tensor ${\cal
R}^{im}$ satisfies
\begin{equation}
{\cal R}^{im} \equiv g_{kn} {\cal R}^{ikmn} = \frac{1}{2}R
g^{im}(3q_1+q_2) + R^{im}(q_2+q_3) \,, \label{susca1}
\end{equation}
and so vanishes in a generic curved spacetime when the non-minimal
coupling parameters are linked through two relations $3q_1{+}q_2=0$ and
$q_2+q_3=0$. Analogously, the scalar ${\cal R}$ given by
\begin{equation}
{\cal R} \equiv g_{im}g_{kn} {\cal R}^{ikmn} = g_{im}{\cal R}^{im} = R
(6q_1+3q_2+q_3) \,, \label{susca}
\end{equation}
has zero value in a generic curved spacetime when
$6q_1{+}3q_2{+}q_3{=}0$.

\subsection{Non-minimal light propagation in a cosmological
context: Optical metric and refraction index, phase and
group velocities, and unlighted epochs}
\label{optmet}

\subsubsection{Optical metric and refraction index}
\label{refindex}

\noindent
{\it The optical metric} -
\label{opticalmetric}
Based on the metric (\ref{FLRW}) and on the symmetries of the
spacetime one obtains that the non-vanishing components of
the non-minimal susceptibility tensor
${\cal R}^{ik}_{ \ \ mn}$, see Eq.~(\ref{susceptibility2}),
have the following form
\begin{equation}\label{R1}
{-} {\cal R}^{1t}_{ \ \ 1t} = {-} {\cal R}^{2t}_{ \ \ 1t} = {-} {\cal
R}^{3t}_{ \ \ 1t} {=} (3q_1 {+} 2q_2 {+} q_3)\frac{\ddot{a}}{a} {+}
(3q_1 {+} q_2) \left(\frac{\dot{a}}{a}\right)^2 \,,
\end{equation}
\begin{equation}\label{R4}
{-} {\cal R}^{12}_{ \ \ 12} = {-} {\cal R}^{13}_{ \ \ 13} = {-} {\cal
R}^{23}_{ \ \ 23} {=} (3q_1 {+} q_2)\frac{\ddot{a}}{a} {+} (3q_1 {+}
2q_2+ q_3) \left(\frac{\dot{a}}{a}\right)^2\,,
\end{equation}
where the indices $1,2,3$ correspond to $x,y,z$, respectively.
In addition,
due to the spacetime isotropy the linear response tensor $C^{ikmn}$
can be rewritten in the standard multiplicative form
\begin{equation}\label{C1}
C^{ikmn} = \frac{1}{2\mu(t)} \left(g^{im}_{*} g^{kn}_{*} -
g^{in}_{*}g^{km}_{*}\right) \,,
\end{equation}
where
\begin{equation}\label{C2}
g^{im}_{*} = n^2(t) \delta^{i}_{t} \delta^{m}_{t} -\frac{1}{a^2(t)}
\left(\delta^{i}_{1}\delta^{m}_{1} + \delta^{i}_{2}\delta^{m}_{2} +
\delta^{i}_{3}\delta^{m}_{3} \right)\,,
\end{equation}
is the so-called associated metric (see \cite{BZ05}), and the
function $n^2(t)$ is defined as
\begin{equation}\label{C3}
n^2(t) \equiv \varepsilon(t) \mu(t) \,,
\end{equation}
with  $\varepsilon(t)$ and $\mu(t)$ given by
\begin{equation}\label{C32}
\varepsilon(t) \equiv 1+
2 {\cal R}^{1t}_{\ \ 1t} \,, \quad
\frac{1}{\mu(t)} \equiv 1+ 2 {\cal R}^{12}_{\ \ 12} \,.
\end{equation}
Taking into account that in a cosmological context the global
velocity four-vector is $U^i= \delta^{i}_{t}$, one can see that
the associated metric (\ref{C2}) is in fact an optical metric
\cite{om3,om4} given by
\begin{equation}\label{C4}
g^{im}_{*} = g^{im} + [n^2(t)-1] U^i U^m \,.
\end{equation}
Its inverse is then
\begin{equation}\label{C4in}
\quad g_{km}^{*} = g_{km}
+ \left[\frac{1}{n^2(t)}-1\right] U_k U_m  \,,
\end{equation}
with
\begin{equation}\label{C4compl}
g^{im}_{*}g_{km}^{*} = \delta^i_k \,,
\end{equation}
holding.
The quantity $n(t)$ is then interpreted as an
effective refraction index due to its correspondence, in
conjunction with the optical metric  $g^{im}_{*}$,
to the effective refraction in the geometrical
optics approximation framework.

Let us see this in detail. In the geometrical optics approximation, the
electromagnetic potential  $A_k$ and the field strength
$F_{kl}$ can be put as follows
\begin{equation}
A_k = \widetilde A_k e^{i \Psi} \,, \quad F_{kl}  = i \left[
p_k \widetilde A_l - p_l \widetilde A_k\right]
e^{i \Psi} \,, \quad p_k = \nabla_k \Psi
\,,\label{AP1}
\end{equation}
for some amplitude $\widetilde A_k$ and phase function $\Psi$.
In the leading order approximation the Maxwell equations (\ref{max1})
reduce to $C^{ikmn} \ p_k \ p_m \widetilde A_n = 0$, i.e.,
\begin{equation}
\left[g^{im}_{*} g^{kn}_{*}
- g^{in}_{*}g^{km}_{*}\right] p_k  p_m  \widetilde A_n = 0 \,.
\label{AP2}
\end{equation}
Using the optical metric property $g^{im}_{*}U_m = n^2 U^i$, the
Landau gauge $U^m \widetilde A_m=0$, and the requirement that the
frequency $\omega \equiv U^m p_m$ is non-vanishing, one can show that
Eq.~(\ref{AP2}) leads to the dispersion equation
\begin{equation}
g^{km}_{*}\ p_k \ p_m = 0 \,.
\label{AP7}
\end{equation}
Thus, the propagation of photons non-minimally coupled to the gravity
field is equivalent to their motion along a null geodesic line in an
effective spacetime with metric (\ref{C4}) (see, e.g., \cite{om3,om4}
for details).  In our setting,
the refraction index $n(t)$ determines the effective
phase velocity of light in our FLRW-type spacetime,
namely,
$V_{{\rm ph}}\equiv \frac{c}{n(t)}$.
Thus, let us consider in more detail the
properties of the quantity $n^2(t)$.

\vskip 0.2cm
\noindent
{\it The expression for the refraction index} -
\label{expr}
The square of the effective refraction index can be obtained from
Eqs.~(\ref{C3})-(\ref{C32}) and is given by
\begin{equation}
n^2(t) = \frac{1 - 2 (3q_1 {+} 2q_2 {+} q_3)\frac{\ddot{a}}{a} -2
(3q_1 {+} q_2) \left(\frac{\dot{a}}{a}\right)^2}{1 - 2(3q_1 {+}
q_2)\frac{\ddot{a}}{a} -2 (3q_1 {+} 2q_2{+} q_3)
\left(\frac{\dot{a}}{a}\right)^2 }\,.
\label{R19}
\end{equation}
Notably, it can be put as a function of the Hubble function $H(t)$ and
the acceleration parameter ${-}q(t)$, defined in
(\ref{R2a})-(\ref{R2b}),
as well as of two effective non-minimal coupling
constants,
$Q_1$ and $Q_2$ given by,
\begin{equation}
Q_1 \equiv - 2(3q_1 {+} 2q_2 {+}
q_3) \,, \quad Q_2 \equiv -2 (3q_1 {+} q_2)  \,.
\label{R3}
\end{equation}
As a function of these quantities Eq.~(\ref{R19}) can be rewritten as
\begin{equation}
n^2(t) = \frac{1 + \left[Q_2 - Q_1 q(t)
\right]H^2(t) }{1 + \left[Q_1 - Q_2 q(t) \right] H^2(t)} \,.
\label{R49}
\end{equation}
For the cosmological models with a de Sitter-type final stage (i.e.,
$-q(t\to \infty) \to 1$), Eq.~(\ref{R49}) yields $n^2
\to 1$.  The signs and the values of the parameters $Q_1$ and $Q_2$
are not yet established, but one can discuss several
phenomenological
possibilities using geometric analogies and physical motivations, see
Appendix.

\subsubsection{Phase velocity and group velocity (or energy transfer
velocity)}
\label{pahseveloc}

The phase velocity of an electromagnetic wave is defined as ($c=1$)
\begin{equation}
V_{{\rm ph}} \equiv \frac{\omega}{k} =
\frac{1}{n(t)} \,.
\label{phaseveloc}
\end{equation}
This definition is standard and follows
directly from the dispersion relation \cite{LLP}.

On the other hand, the definition of group velocity, or energy
transfer velocity, $V_{*}$, is connected with the definition of the
electromagnetic field energy flux, and this problem requires a
preliminary discussion. First of all, one should state that formally
the propagation of electromagnetic waves takes place in isotropic
backgrounds with a refraction index equal to one, but under the
influence of a non-minimal coupling the interaction can be
reformulated in terms of an effective refraction index $n(t)$, which
depends on time through the Riemann tensor components. Since, as we
have seen, the non-minimal coupling of photons to gravity is
equivalent to the consideration of some isotropic dispersive medium,
it seems reasonable to take the relevant stress-energy tensor of the
electromagnetic field for the definition of the energy transfer
velocity $V_{*}$. Now, there is a number of definitions for the
electromagnetic stress-energy tensor in a medium, the best known are
the ones by Minkowski \cite{Minkowski}, Abraham
\cite{Abraham1,Abraham2}, Grot-Eringen-Israel-Maugin
\cite{GE1,GE2,IsraelT,Maugin78}, Hehl-Obukhov \cite{OH}, and de
Groot-Suttorp \cite{GS}. The energy flux four-vectors in these
definitions differ from one another. Thus for us a choice between the
several definitions is considered as an ansatz.

We follow the Hehl-Obukhov definition \cite{OH}, according to which
the stress-energy tensor formally is as in vacuum,
$T^{\rm electromag}_{ik}$,
i.e.,
$T^{\rm electromag}_{ik} \equiv \frac{1}{4} g_{ik} F^{mn}
F_{mn} - F_{im} F_{k}^{ \ m}$,
and thus is symmetric, traceless and does not depend on
the macroscopic velocity of the medium.  Nevertheless, the
ponderomotive force
\begin{equation} F^l  =
\nabla_k T^{{\rm electromag}\,kl}= F^{l}_{ \ m}
\nabla_k (H^{km}-F^{km}) \label{THO1}
\end{equation}
is non-vanishing. In contrast to the vacuum case, this stress-energy
tensor is not a conserved quantity, its four-divergence is equal to
the ponderomotive force.
According to this definition \cite{OH}, the energy flux $I^i$ is
\begin{equation}
I^i \equiv (g^{ip} - U^i U^p) T^{\rm electromag}_{pq} U^q = \eta^{ipq}
B_p E_q \,, \label{THO2}
\end{equation}
where
\begin{equation}
\eta^{ipq} \equiv \epsilon^{ipqs} U_s \,, \quad B_p \equiv F^{*}_{pj}
U^j \,, \quad E_q \equiv F_{qj} U^j \,,
\label{THO2-1}
\end{equation}
and the energy-density scalar $W$ is
\begin{equation}
W \equiv U^p T^{\rm electromag}_{pq} U^q = - \frac{1}{2} (E^m E_m + B^m
B_m) \,.
\label{THO3}
\end{equation}
When we deal with test electromagnetic waves in an isotropic universe,
the necessary quantities can be found as follows. The non-minimally
extended Maxwell equation can be rewritten now in the form
\begin{equation}
\nabla_k \left[\frac{1}{\mu} F^{ik} + \frac{n^2-1}{\mu} (F^i_{\ m} U^k
- F^k_{\ m} U^i)U^m \right] = 0 \,,
\label{Meq}
\end{equation}
where $\mu(t)$ and $n^2(t)$ are given by (\ref{C3})-(\ref{C32}).
Let the
direction along which the electromagnetic wave propagates be
$x^1$, say.
Then Eq.~(\ref{Meq}) admits the following solution for the
potential four-vector,
\begin{equation}
A_i(t,x^1) = \delta_i^2 \ b(t) \cos{\Psi} \,, \label{Meq2}
\end{equation}
with
\begin{equation}
\Psi = \Psi(t_0) +
k_1[x^1 - f(t)] \,,
\label{Meq2-1}
\end{equation}
and where $k_1$ is a  constant wave-vector,
$b(t)$ the amplitude factor, and
$f(t)$ the
frequency function. These satisfy the following equations,
\begin{equation}
\frac{df}{dt} =
\frac{1}{an}\sqrt{1 + \left(\frac{a \mu}{b k_1^2}\right) \
\frac{d}{dt}\left[\left(\frac{a n^2 }{\mu}\right) \frac{db}{dt}
\right] }\,,
\label{Meq3}
\end{equation}
\begin{equation}
\frac{d}{dt} \left[\left(\frac{b^2 a
n^2}{\mu}\right) \frac{df}{dt} \right] = 0 \,.
\label{Meq3-1}
\end{equation}
In the geometrical optics approximation, i.e., for
short waves, one has ($k_1 \to
\infty$) and (\ref{Meq3}) gives $\dot{f} =
\frac{1}{an}$, i.e., the quantity $\omega(t) \equiv k_1\dot{f}=
\frac{k_1}{an}$ plays the role of a time dependent frequency, the
quantity $k(t) \equiv \frac{k_1}{a}$ is a wave-vector modulus, and
the dispersion relation takes the form $\omega(t) {=}
\frac{k(t)}{n(t)}$. In this approximation, Eq.~(\ref{Meq3-1})
gives $b(t) = b_0
\sqrt{\frac{\mu(t)}{n(t)}}$.

From a physical point of view (see, e.g., \cite{LLP}) it is reasonable
to calculate the non-vanishing energy flux four-vector component and
the energy density scalar averaged over a wave period, in which case
one can use $\langle\cos^2\Psi\rangle = \langle\sin^2\Psi\rangle =
\frac{1}{2} $ and $\langle\cos{\Psi} \sin{\Psi}\rangle =0$, where
$\langle\quad\rangle$ denotes average over a period. In the geometrical
optics approximation the energy flux $I^i$ and the energy-density scalar
$W$ have the form
\begin{equation}
\langle I^1 \rangle = - \frac{1}{a^4} \langle F_{t2} F_{12} \rangle =
\frac{b^2 k_1^2}{2na^5} \,,
\label{meq7}
\end{equation}
and
\begin{equation}
\langle W \rangle = \frac{1}{2a^2}
\langle F^2_{t2} \rangle + \frac{1}{2a^4} \langle F^2_{12} \rangle =
\frac{b^2 k_1^2 (n^2+1)}{4a^4 n^2} \,.
\label{meq7-2}
\end{equation}
Thus, the physical component of the energy transfer velocity $V_{*}$
is
\begin{equation}
V_{*} = \frac{\sqrt{- g_{11} \langle I^1 \rangle \langle I^1
\rangle}}{\langle W \rangle} =
\frac{2n}{n^2+1} \,.
\label{Meq8-0}
\end{equation}
The function $V_{*}(n(t)) = \frac{2n}{n^2+1}$ is appropriate for
the description of the energy transfer velocity, since depending on the
refraction index $n$ it vanishes in the limits $n \to 0$ and $n \to
\infty$. In addition,
its value is maximum in pure vacuum since
$V_{*} = 1$ when $n=1$, and generically satisfies the inequality
$|V_{*}(n(t))| \leq 1$. Below, to simplify terminology, we indicate
this energy transfer velocity as group velocity $V_{\rm gr}$, i.e.,
$V_{\rm gr} \equiv V_{*}$, and so
\begin{equation}
V_{\rm gr} = \frac{2n}{n^2+1} \,.
\label{Meq8}
\end{equation}

\subsubsection{Unlighted cosmological epochs: definition}
\label{unlightepochssub}

It is remarkable that
the square of the effective refraction index can take negative values,
$n^2(t)<0$, for some set of the parameters $Q_1$ and $Q_2$.  We call
the time intervals for which $n^2(t)<0$ as unlighted epochs, since during
these periods of time the refraction index is a pure imaginary
quantity, and the phase and group velocities of the electromagnetic
waves are not defined. The function $n^2(t)$ can change sign at the
moments $t_{(s)}$
of the cosmological time when the numerator or the
denominator in Eq.~(\ref{R49}) vanish.  When the numerator vanishes,
one has $n(t_{(s)})=0$, $V_{{\rm ph}}(t_{(s)})=\infty$, and $V_{{\rm
gr}}(t_{(s)})= 0$. When the denominator vanishes, one has
$n(t_{(s)})=\infty$, $V_{{\rm ph}}(t_{(s)})=0$, and $V_{{\rm
gr}}(t_{(s)})= 0$. In both cases the unlighted epochs appear and disappear
when the group velocity of the electromagnetic waves vanishes, i.e.,
at these points there is no energy transfer.
In our terminology the times $t_{(s)}$ are
the unlighted epochs boundary points.
For this reason, the condition $V_{{\rm
gr}}(t_{(s)})= 0$ sets the criterion for the unlighted epoch appearance
or disappearance. In other words, the unlighted epochs start and finish,
when the associated metric, i.e., the optical
metric, given in Eq.~(\ref{C2}) becomes singular.

In our cosmological context we distinguish
three types of unlighted epochs:
\begin{description}
\item 1. Unlighted epochs of the first type: These
epochs start at $t_{(1)}{=}0$ with
$n^2(0)<0$ and finish with $n^2(t_{(2)}){=}0$.
\item 2. Unlighted epochs of the second type: These
epochs start at $t_{(1)}>0$ with
$n^2(t_{(1)}){=}0$ and finish at $t_{(2)}>t_{(1)}$ with
$n^2(t_{(2)}){=}0$.
\item 3. Unlighted epochs of the third type: These
epochs appear when at least one
boundary point has a refraction index
with an infinite value, i.e.,
$n^2(t_{(1)}){=}\infty$ or $n^2(t_{(2)}){=}\infty$.
Of course, in this type both
quantities may be infinite.
\end{description}
Clearly, only one unlighted epoch of the first
type can exist, whereas the number of unlighted epochs of the second and
third
types is predetermined by the guiding parameters $Q_1$ and $Q_2$.

\section{Unlighted cosmological epochs: analytical study}
\label{analyticalstudy}

\subsection{The cosmology}
\label{choiceag}


In order to illustrate analytically the physics of unlighted epochs,
let us consider a scale factor $a(t)$ given by a
stretched exponential function, namely,
\begin{equation}
a(t)=a_0 \exp\{ ({\cal H}t )^{\nu}\}\,,
\label{Kohl1}
\end{equation}
for some constant $\cal H$ and exponent $\nu$.
This function was introduced by Kohlrausch \cite{Kohl1} in 1854 and
now is systematically used in various physical and mathematical
contexts (see, e.g., papers concerning the applications of the
Kohlrausch-Williams-Watts (KWW) function
\cite{Kohl2,Kohl4_1}).  When
$\nu {=}1$ the function (\ref{Kohl1}) coincides with the standard de
Sitter exponent. When $\nu{=}2$ one deals with an anti-Gaussian
function obtained in \cite{Arch1} as an exact solution of
a model
with Archimedean-type interaction between dark energy and dark
matter. This stretched exponent also appeared in \cite{6}
in the context of generalized Chaplygin
gas models. For the function (\ref{Kohl1}) one obtains from
Eqs.~(\ref{R2a})-(\ref{doth}),
\begin{equation}
H(t)= \nu {\cal H}^{\nu} t^{\nu-1} \,, \quad -q(t)= 1 +
\frac{\nu{-}1}{\nu} ({\cal H}t )^{-\nu}\,, \quad \dot{H}(t) = \nu
(\nu{-}1) {\cal H}^{\nu} t^{\nu-2}\,.
\label{Kohl2}
\end{equation}
It is reasonable to assume that $\nu$ is positive in order to
guarantee that $a(t \to \infty) \to \infty$ and $-q(t \to \infty) \to
1$. Clearly, when $0<\nu<1$, $\dot{H}$ is negative, and
the Hubble function $H(t)$ vanishes at $t
\to \infty$. Our ansatz is that $\nu\leq2$. Note that
for $\nu<2$ one has
$\dot{H}(t \to \infty) \to 0$, and thus, $n^2(t \to \infty) \to
1$. Let us focus on the properties of the function $n^2(t)$ for a
scale factor of the Kohlrausch type Eq.~(\ref{Kohl1}),
\begin{equation}
n^2(t)= \frac{({\cal H}t )^{2-\nu}+\nu^2 {\cal H}^{2} (Q_1{+}Q_2)
({\cal H}t)^{\nu}+ \nu (\nu{-}1){\cal H}^{2}Q_1}{({\cal H}t
)^{2-\nu}+\nu^2 {\cal H}^{2} (Q_1{+}Q_2) ({\cal H}t)^{\nu}+ \nu
(\nu{-}1){\cal H}^{2}Q_2}\,.  \label{Kohl3}
\end{equation}
When $\nu=1$, there are no unlighted epochs, since $n^2(t) \equiv
1$. When $\nu=2$, both the numerator and the denominator in
(\ref{Kohl3}) are quadratic functions of $t$, which means that
depending on
the values of the parameters $Q_1$ and $Q_2$,
one can find zero, one, two, three or
four boundary points $t_{(s)}$. In other words, the
number of unlighted epochs can be zero, one or two. 
These models represent perpetually accelerated
universes since$-q(t)$ is non-negative for $t \geq
0$. When $\nu$ is a
perfect rational  $\nu {=} \frac{m_1}{m_2}$ with $m_1$
and $m_2$ natural numbers and $m_1<m_2$,
the numerator and denominator of the function
(\ref{Kohl3}) can be rewritten as a polynomial of order
$2m_2{-}m_1$. Such a polynomial has $2m_2{-}m_1$ roots, part of them,
say $k$, can be real and positive defining the corresponding number of
unlighted epochs.

\begin{figure}
[t]
\centerline{\includegraphics[width=11.00cm,height=6.75cm]{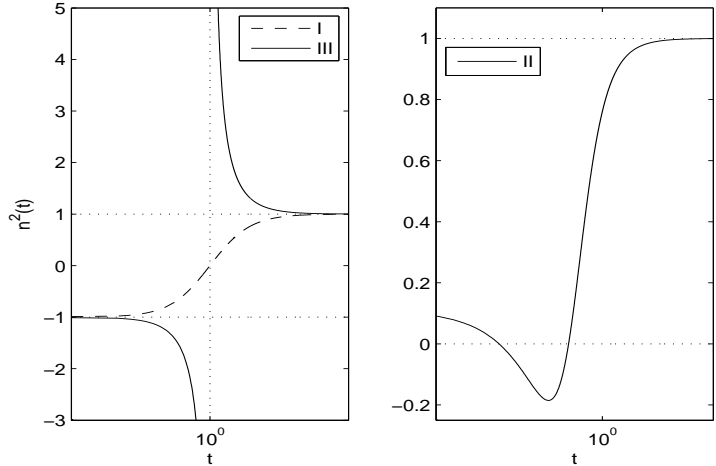}}
\caption {{\small A plot of $n^2(t)$ as a function of time $t$ in the
Kohlrausch-type model.  Left panel: Contains sketches of
unlighted epochs of the first and third types. For both types the
Kohlrausch parameter is set to $\nu =1.00001$.
The non-minimal parameters are the
following: $Q_2=-Q_1=10000$ for the model of the first type, and
$Q_2=-Q_1=-10000$ for the model of the third type. The unlighted epoch of
the first type appears in the plot as a dashed curve.
It is continuous, it starts
with a negative value, $n^2(0)={-}1$, and has one zero
(the point in which the phase velocity is infinite and the group
velocity is vanishing). The unlighted epoch of the third type
appears in the plot as a solid curve. It is
discontinuous, the function $n^2(t)$ has one infinite jump
(in this point the phase and group velocities are vanishing).
Right panel: It is illustrated a
unlighted epoch of the second type with
parameters $\nu=2/3$, $Q_1=-81/800$, $Q_2=-729/800$. There are two
points, in which the refraction index and group velocity vanish and
the phase velocity becomes infinite. }}
\label{kol}
\end{figure}

\subsection{The choice for the non-minimal parameters}
\label{choicesub}

In order to develop the unlighted epochs
in the cosmological models of interest one has to choose the
non-minimal parameters. Here
we give two example which are enough to have a
feeling of the physics. In one case
we put $Q_1=-Q_2$, and define $Q\equiv Q_1$.
In the other case we choose
$Q_1=-\frac{81}{800 {\cal
H}^{2}}$ and $Q_2=-\frac{729}{800 {\cal H}^{2}}$.

\subsection{Two cosmological models and their unlighted epochs}
\label{2m}

\subsubsection{Universe
with $\nu<2$ and $Q_1=-Q_2\equiv Q$}

We now put $Q_1=-Q_2$ and define $Q=Q_1$. Then,
the square of the refraction index in Eq.~(\ref{Kohl3})
has now the form
\begin{equation}
n^2(t)= \frac{({\cal H}t )^{2{-}\nu}+ \nu (\nu{-}1){\cal
H}^{2}Q}{({\cal H}t )^{2{-}\nu}- \nu (\nu{-}1){\cal H}^{2}Q}\,.
\label{Kohl4}
\end{equation}
There is now only one time moment $t=t_{(1)}$ when $n^2$ is either
equal to zero or infinite. Equivalently, there is only one moment when
the group velocity takes zero value.  When $(\nu{-}1)Q<0$, the
denominator is positive, $n^2(0)={-}1<0$, and the numerator vanishes
at $t_{(1)}= \frac{1}{{\cal H}}\left[\nu {\cal H}^{2}|(\nu{-}1) Q|
\right]^{\frac{1}{2-\nu}}$; clearly, we deal with an unlighted epoch
of the first type, see the dashed line in Fig.~\ref{kol} left panel.  When
$(\nu{-}1)Q>0$, the numerator is positive, $n^2(0)={-}1<0$, and the
denominator vanishes at $t_{(1)}= \frac{1}{{\cal H}}\left[\nu
(\nu{-}1){\cal H}^{2}Q \right]^{\frac{1}{2-\nu}}$; clearly, we deal
with an unlighted epoch of the third type with $n^2(t_{(1)}) =\infty$,
see the solid line in Fig.~\ref{kol} left panel.

\subsubsection{Universe
with $\nu = \frac{2}{3}$ and $Q_1=-\frac{81}{800 {\cal
H}^{2}}$, $Q_2=-\frac{729}{800 {\cal H}^{2}}$}

Using the auxiliary variable $x \equiv (27 {\cal H}t )^{\frac{2}{3}}$
we can rewrite the square of the refraction index in Eq.~(\ref{Kohl3})
in the following
form
\begin{equation}
n^2(t)= \frac{x^2+ 4 {\cal H}^{2} (Q_1{+}Q_2) x - 18 {\cal
H}^{2}Q_1}{x^2 + 4 {\cal H}^{2} (Q_1{+}Q_2) x - 18 {\cal H}^{2}Q_2}\,.
\label{Kohl5}
\end{equation}
It is easy to find general conditions, when the denominator has no
roots and the numerator has two positive roots. For instance,
Fig.~\ref{kol} right panel,
illustrates the case with $Q_1=-\frac{81}{800 {\cal
H}^{2}}$ and $Q_2=-\frac{729}{800 {\cal H}^{2}}$. We deal now with an
unlighted epoch of the second type, which starts  at
$t_{(1)}{=} \frac{1}{{\cal H}}\left[\frac{3}{40}(3 - \sqrt{5})
\right]^{\frac{3}{2}}$, and finishes at
$t_{(2)}{=} \frac{1}{{\cal H}}\left[\frac{3}{40}(3 + \sqrt{5})
\right]^{\frac{3}{2}}$.

\section{Unlighted cosmological epochs: Numerical study. The
refraction index, phase and group velocities for six models and the
effective photon pathlength}
\label{num}

\subsection{The cosmology}
\label{choice}

In \cite{Arch1,Arch2} several cosmological models were presented
through numerical results. Here we study non-minimal light propagation
in six such models. They include, perpetually accelerated universes,
periodic universes, universes with one transition, two transition and
three transition points, and quasiperiodic universes.

\subsection{The choice for the non-minimal parameters}
\label{choicesub2}

To study non-minimal light propagation
in the universes indicated above, there is a plethora
of possibilities to make a choice of the non-minimal parameters.
A sample of those possibilities is presented in the Appendix.
For our study we make one choice, bearing in mind that the
other choices will reproduce qualitatively the same
results. We choose that non-minimal susceptibility scalar
${\cal R}$ vanishes, ${\cal R}=0$.
From Eq.~(\ref{susca})  this means
$6q_1{+}3q_2{+}q_3=0$ and thus, from Eq.~(\ref{R3}),
one has $Q_1+Q_2=0$. Putting $Q\equiv
Q_1=-Q_2$, one obtains in this one-parameter example
the following expression for $n^2(t)$,
\begin{equation}
n^2(t) = \frac{1 - Q H^2(t)\left[1+ q(t)\right]}{1 +
Q H^2(t)\left[1+ q(t)\right]} = \frac{1 + Q \dot{H}(t)}{1 - Q
\dot{H}(t)}\,.
\label{R43}
\end{equation}
Now that we see an explicit expression for $n^2(t)$
our choice can be motivated as follows. First, it is an example
that best illustrates the several unlighted epochs. This is
because only one function, $\dot{H}$, guides the
behavior of the effective refraction index. Second, when $t \to
\infty$, $\dot{H} \to 0$ and thus $n^2 \to 1$ providing $V_{\rm ph} \to
V_{\rm gr} \to 1$ in our late-time Universe, as it should be. We will
consider this one-parameter example as the one which will provide a
substratum for our numerical analysis.

\subsection{Six cosmological models, their unlighted epochs,
and the photon pathlength}
\label{6m}

The results of our numerical calculations for
the six basic cosmological
models in which an Archimedean-type interaction between dark energy
and dark matter plays a main role are presented in
Figs.~\ref{pepau}--\ref{evolutionarymfig}.  The panels $({\rm a})$
of the figures
display the plots of $\dot{H}(t)$.  The panels $({\rm b})$ display the
plots of the square of the refraction index $n^2(\tau)$, where $\tau$
is a logarithmic time defined by $\tau \equiv \log{x}$, with $x\equiv
{\frac{a(t)}{a(t_0)}}$.  For the calculations we use formula
(\ref{R43}) putting there the values of the guiding parameter $Q$ for
which the square of the refraction index $n^2(t)$ can be negative for
some period of time.  The panels $({\rm c})$ display the plots of the
electromagnetic wave phase velocity. Here we use values of $Q$, for
which $n^2(t)>0$ and the phase velocity happens to be a pure real
function. The panels $({\rm d})$ display the plots of the group
velocity for the same values of the guiding parameters $Q$.
Additional internal windows clarify fine details of the plots.

\subsubsection{Perpetually accelerated universe}
\label{perpet}

\begin{figure}
[b]
\centerline{\includegraphics[width=11.00cm,height=6.55cm]{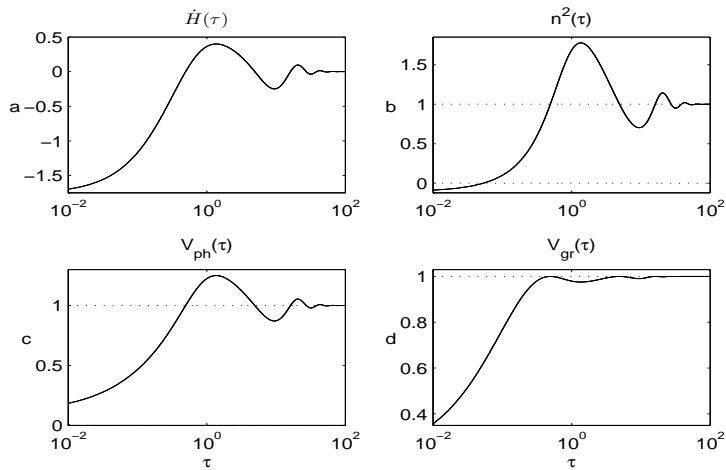}}
\caption {{\small Perpetually accelerated universe. The plots
in panels (a), (b), (c), and (d)
are presented for a typical model with the
following parameters: $\xi{=}0.35$, $\sigma {=}{-}0.99$ (i.e.,
$3\xi{+}\sigma \simeq 0.06>0$), ${\cal V}_{(0)} {=}1$, $E_{(0)} {=}
0.0205$, $\lambda_{(0)} {=}1$, $\rho_{*} {=}0.333 \cdot 10^{-4}$, and
$\Pi^{\prime}(1){=}{-}1$.
Panel (a): illustrates the behavior of the function $\dot{H}(t)$, the
plot of which has a finite number of visible damped oscillations, and
tends to zero asymptotically; it relates to the case of the
non-negative acceleration parameter ${-}q(t)$, i.e., there are no
transition points in this model, there is no partition of the universe
history into epochs, but there are a few eras, the start and finish of
which are marked by the extrema of the function $\dot{H}$.
Panel (b): The plot of
the square of the refraction index $n^2(\tau)$ is presented in the
panel (b) for the parameter $Q=0.7$. It contains at least three eras,
in which the refraction index exceeds one, three eras with $n<1$, and
$n^2(\tau)$
tends asymptotically to one at $t \to \infty$. There is an unlighted epoch
of the first kind, which is characterized by negative values of the
function $n^2(t)$; it starts at $t=0$ and extends to the middle of the
first era. At the end of this unlighted epoch the refraction index takes
zero value, and starting from this point electromagnetic waves
can propagate and transfer information into the
universe.
Panel (c): This panel displays the phase velocity of the
electromagnetic waves as a function of time; the plot relates to a
parameter $Q=-0.55$; this choice guarantees that $n^2(t)$ is positive,
and $V_{{\rm ph}}$ is a real function everywhere. The plot of the
phase velocity reflects the history of universe, i.e., it has extrema
just at the moments when one era is changing into another. Clearly, the
waves move more slowly in the early universe, when the curvature is
large, then there are few eras with oscillations of the phase velocity
near the vacuum speed of light, and finally, $V_{{\rm ph}}$ tends
to one asymptotically.
Panel (d): The last panel shows the behavior of
the group velocity of the electromagnetic waves; the calculations are
made for the same values of the guiding non-minimal
parameters as for the phase
velocity. Clearly, $V_{{\rm gr}}$ does not exceed one, it tends to
one asymptotically, and an energy transfer takes place slowly in the
early universe.
}}
\label{pepau}
\end{figure}

The first class of models is the class of perpetually accelerated
universes.  This class arises when $-q(t)$ is non-negative for $t \geq
0$ and so there are no points in which $-q(t)$ could change
sign. In this sense the history of such class of universes includes
only one epoch.

In this model the dark energy energy density $\rho$
is always non-negative and the dark energy $\Pi$ is non-positive, see
\cite{Arch2} for more details.
Although there is only one epoch for these universes, one can divide
this epoch into eras. The
extrema of the functions $-q(t)$, $H(t)$, $\rho(t)$, and $\Pi(t)$
give the boundary points between the eras, see
\cite{Arch2} for more details.
We do not plot $-q(t)$, $H(t)$, $\rho(t)$, and $\Pi(t)$, see
\cite{Arch2} for such plots.

We are interested in light
propagation and unlighted epochs. So we plot $\dot{H}(t)$,
the square of the refraction index $n^2(t)$,  the
phase velocity $V_{{\rm ph}}(t)$, and for the group velocity $V_{{\rm
gr}}(t)$. Fig.~\ref{pepau} shows that indeed for the particular
case chosen there are at least six eras which appear visually.

The first era is an era of superacceleration.  During this era the
function $\dot{H}(t)$ grows monotonically and reaches its global
maximum, as it is shown in the panel (a) of Fig.~\ref{pepau}. The same
behavior happens for the square of the refraction index $n^2$, for the
phase velocity $V_{{\rm ph}}(t)$, and for the group velocity $V_{{\rm
gr}}(t)$, as it is shown in panels (b), (c), and (d) of
Fig.~\ref{pepau}, respectively.

The second, third, and following eras appear one after the other as
the parameters $-q(t)$, $H(t)$, $\rho(t)$, and $\Pi(t)$ of the
universe relax to a state with asymptotically constant positive
values, namely, $-q_{\infty}$, $H_{\infty}$, $\rho_{\infty}$ and
negative $\Pi_{\infty}$.  For the parameters that guide the properties
of light propagation we see from Fig.~\ref{pepau} that $\dot{H}(t)$
asymptotically vanishes, see panel (a), and the refraction index, the
phase velocity and the group velocity tend quasiperiodically
asymptotically to one, see panel (b), (c), and (d), respectively.

Other features can be mentioned: (i) There are at least three eras, in
which the refraction index exceeds one, $n>1$, and three eras with
$n<1$.  (ii) There is one unlighted epoch of the first type, which is
characterized by negative values of the function $n^2(t)$. This unlighted
epoch starts at $t=0$ with a non-zero value of $n^2$, and extends up
to the middle of the first era. At the end of this unlighted epoch the
refraction index takes zero value (the corresponding effective phase
velocity would be infinite), and starting from this point the
electromagnetic waves can propagate and transfer information
into the universe. (iii) Clearly, the waves move more slowly in the
early universe, when the curvature is large. (iv) Let us note that the
plot of the phase velocity properly reflects the history of the
universe, i.e., it has extrema just at the moments when one era is
changing into another.

\subsubsection{Periodic universe}
\label{peridicu}
\begin{figure}
[b]
\centerline{\includegraphics[width=11.00cm,height=6.55cm]{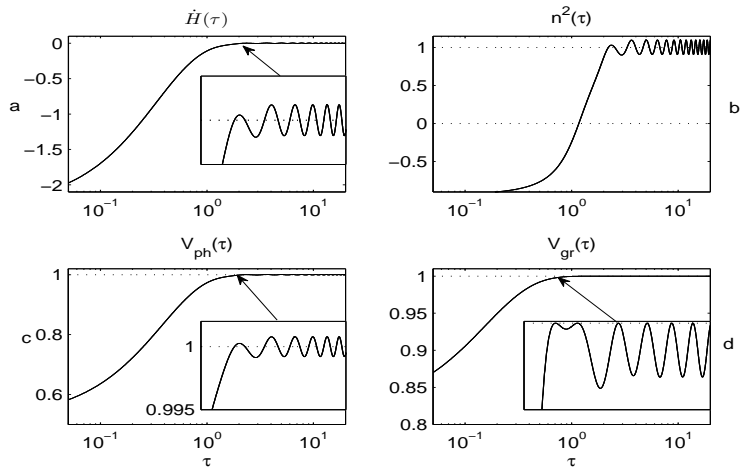}}
\caption {{\small Periodic universe.
The plots
in panels (a), (b), (c), and (d)
are presented for a typical model with the
following parameters: $\xi{=}0.1$, $\sigma
{=}{-}0.299999$ (i.e., $3\xi{+}\sigma \simeq 10^{-6}>0$), ${\cal
V}_{(0)} {=}1$, $E_{(0)} {=} 0.0205$, $\lambda_{(0)} {=}1$, $\rho_{*}
{=}0.333 \cdot 10^{-4}$, and $\Pi^{\prime}(1){=} 0.01$.
The model is characterized
by infinite number of transition points, in which the acceleration
parameter ${-}q$ changes sign.
Panel (a): The function $\dot{H}(t)$ (for $Q=15$) grows monotonically in
the early universe and
then oscillates near zero value.
Panel (b):
The plot of the function $n^2(t)$ displays, first, an unlighted epoch of
the first
type,
then grows monotonically up to the end of the first era, and then
oscillates near the value $n=1$.
Panel (c):
This is a  plot of the phase velocity of
the electromagnetic waves (for $Q=-0.25$). The curve follows
the plot of $\dot{H}$, indicating the starting and finishing points of
the corresponding epochs in the history of the universe.
Panel (d):
This is a  plot of
the group velocity. It shows it also grows monotonically during
the first era, then starts to oscillate so that its maximal value
reaches the speed of light in vacuum.
}}
\label{periou}
\end{figure}

\begin{figure}
[b]
\centerline{\includegraphics[width=11.00cm,height=6.55cm]{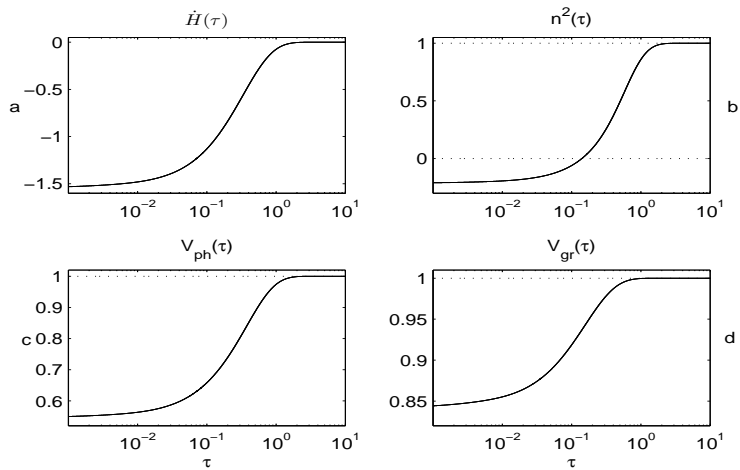}}
\caption {{\small One transition point universe.
This is an example of a universe with one transition point
in its evolution.  The plots in panels (a), (b), (c), and (d) are
presented for a typical model with the following parameters:
$\xi{=}0.1$, $\sigma {=}50$, ${\cal V}_{(0)} {=}1$, $E_{(0)} {=}
0.0205$, $\lambda_{(0)} {=}1$, $\rho_{*} {=}0.333 \cdot 10^{-4}$, and
$\Pi^{\prime}(1){=} {-} 5$.  The plots of all functions presented in
the panels (a), (b), (c) and (d), look like the plot for the hyperbolic
tangent. The plots have
no extrema and illustrate the one-fold transition from
a universe expanding with negative acceleration to a universe with
accelerated expansion. The history of this universe is clearly
divided into two epochs of acceleration/deceleration without
distinguished eras inside. This is clearly seen in
panel (a) for $\dot{H}$.
During the first epoch of deceleration an
unlighted epoch can arise, see panel (b) for which
$Q=1$ was chosen,
and both phase and group velocities are less than the speed of
light in vacuum. The second epoch can be characterized by
parameters which are close to cosmological values
measured nowadays. In particular, the phase
and group velocities during all the second epoch, see panels (c) and
(d) where $Q=-0.35$ was chosen, are close to one.
}}
\label{onetranspoin}
\end{figure}
\begin{figure}
[t]
\centerline{\includegraphics[width=11.000cm,height=6.55cm]{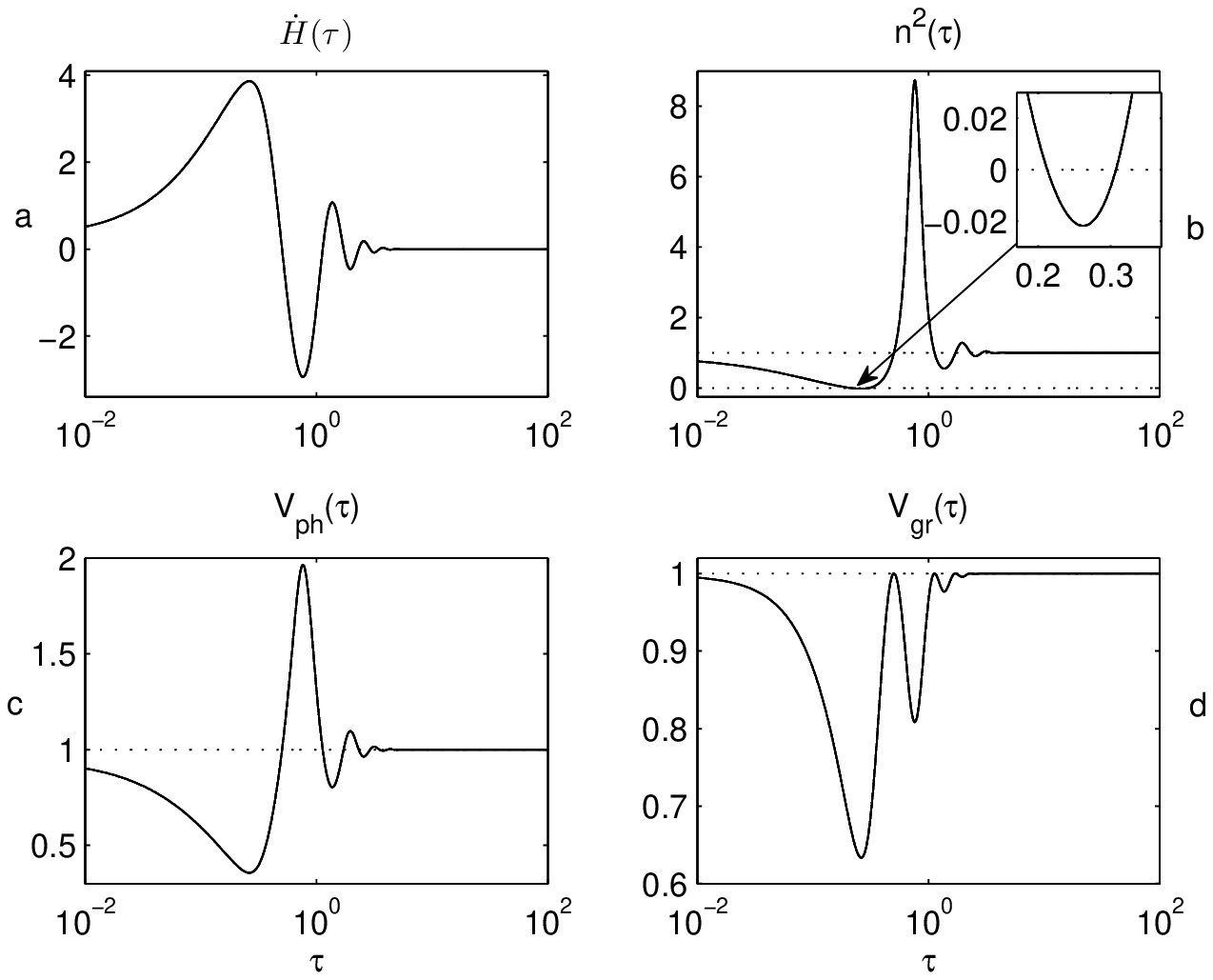}}
\caption {{\small Two transition points universe.
This gives an example of a universe evolution with two
transition points and one extremal point.
The plots in panels (a), (b),
(c), and (d) are presented for a typical model with the following
parameters: $\xi{=}0.1$, $\sigma {=}{-}0.08$, ${\cal V}_{(0)} {=}1$,
$E_{(0)} {=} 0.0205$, $\lambda_{(0)} {=}1$, $\rho_{*} {=}0.333 \cdot
10^{-4}$, and $\Pi^{\prime}(1){=} {-}15$.  Panel (a)
displays the function $\dot{H}$.
In panel (b)
one can see
that the plot for the square
the refraction index $n^2(t)$ contains an unlighted epoch of the
second type since the function $n^2(t)$, here for
$Q=-0.28$, vanishes and then
takes negative values at the end of the first era of the first epoch,
in contrast to the case, when an unlighted epoch appears at $t=0$.
In panel (c) the phase velocity is plotted.  Other new
feature is that in the early universe
the group velocity, calculated here for $Q=0.2$, is close
to the vacuum speed of light, and then this velocity reaches
the same value asymptotically at $t \to \infty$, see panel (d).
}}
\label{twotranspoin}
\end{figure}

The second class of models is the class of periodic universes. This
class arises when the equation $q(t)=0$ has an infinite number of
roots, and the history of the universe splits into an infinite number
of identical epochs with accelerated and decelerated expansions.

The acceleration parameter $-q(t)$, and the Hubble function $H(t)$,
oscillate with fixed frequency and amplitude after the second
transition point \cite{Arch2}. These oscillations are reflected in the
oscillations of the energy density and pressure of the dark energy.
We do not plot $-q(t)$, $H(t)$, $\rho(t)$, and $\Pi(t)$, see
\cite{Arch2} for such plots.

We are interested in light propagation and unlighted epochs. From the
plots
Fig.~\ref{periou} we see that $\dot{H}(t)$ also oscillates, which in
turn reflects again in the behavior of the refraction index $n^2(t)$,
the phase velocity $V_{{\rm ph}}(t)$, and the group velocity
$V_{{\rm gr}}(t)$.

Other features can be mentioned: (i) The plot of the function
$n^2(t)$, see panel (b) of Fig.~\ref{periou}, displays an unlighted
epoch of the first kind, which extends from $t{=}0$ into the middle of
the first era.

\subsubsection{One transition point universe}
\label{univevol}
The third class of models is the class of universes with one
transition point during their evolution.  This class arises when the
acceleration parameter $-q(t)$ is a deformed Heaviside step-function,
or a hyperbolic tangent (see \cite{Arch2}).

The acceleration parameter $-q(t)$ has the property that the change of
a deceleration epoch into an acceleration epoch takes place only once
in a narrow period of time, and the plot looks like a typical plot for
a phase transition. The second epoch, characterized by an accelerated
expansion, looks like the de Sitter-type stage with $\rho{+}\Pi=0$ and
vanishing $E$ and $P$. The first and second epochs are not divided
into eras within this model.  We do not plot $-q(t)$, $H(t)$,
$\rho(t)$, and $\Pi(t)$, see \cite{Arch2} for such plots.
We are interested in light propagation and unlighted epochs.
Fig.~\ref{onetranspoin} illustrates this class of models.

\subsubsection{Two transition points universe}
\label{notrans}
The fourth class of models is the class of universes with two
transition points and one extremal point during their evolution.  This
class arises when the acceleration parameter $-q(t)$ has a double
root, i.e., an extremal point, inside the second epoch of
acceleration. This model is also
distinguished by the fact that the dark matter pressure reaches a
maximum at the end of the first era of the first acceleration epoch.
We do not plot $-q(t)$, $H(t)$, $\rho(t)$, and $\Pi(t)$, see
\cite{Arch2} for such plots.

We are interested in light propagation and unlighted epochs.
Fig.~\ref{twotranspoin} illustrates this class of models.  The
sophisticated behavior that appears in this model is reflected
explicitly in the form of the plot $\dot{H}(t)$, see
Fig.~\ref{twotranspoin}.  In this scenario there is an unlighted epoch
of the second kind, where the function $n^2(t)$ vanishes at some $t$,
with $t>0$, then takes negative values at the end of the first era of
the first epoch, and then vanishes again. Thus this unlighted epoch is
separated by two points, where the refraction index vanishes and the
effective phase velocity takes infinite values.

\begin{figure}
[h]
\centerline{\includegraphics[width=11.000cm,height=6.55cm]{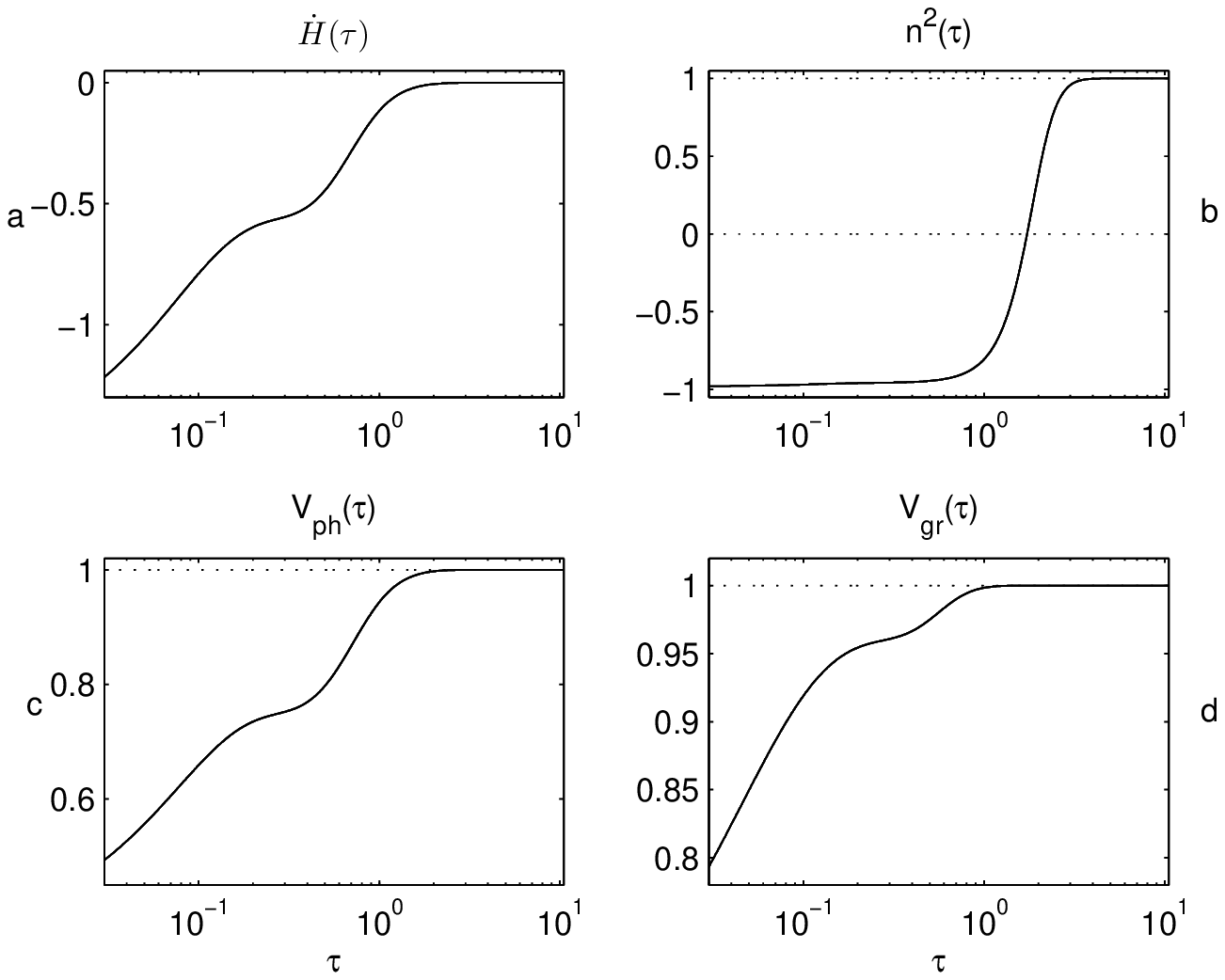}}
\caption {{\small Three transition points universe.  The parameters of
the model are the following: $Q= 80$ for panels (a) and (b),
$Q=-0.5$ for panels (c) and (d), $\xi{=}0.1$, $\sigma {=}1$,
${\cal V}_{(0)} {=}1$, $E_{(0)} {=} 0.0205$, $\lambda_{(0)} {=}1$,
$\rho_{*} {=}0.333 \cdot 10^{-4}$, and $\Pi^{\prime}(1){=} {-}5$.  The
universe passes through two epochs of deceleration and two epochs of
accelerated expansion.  The unlighted epoch harbors now the first and
second epochs of deceleration, as well as the first epoch of
accelerated expansion.  }}
\label{threetranspoin}
\centerline{\includegraphics[width=11.00cm,height=6.55cm]{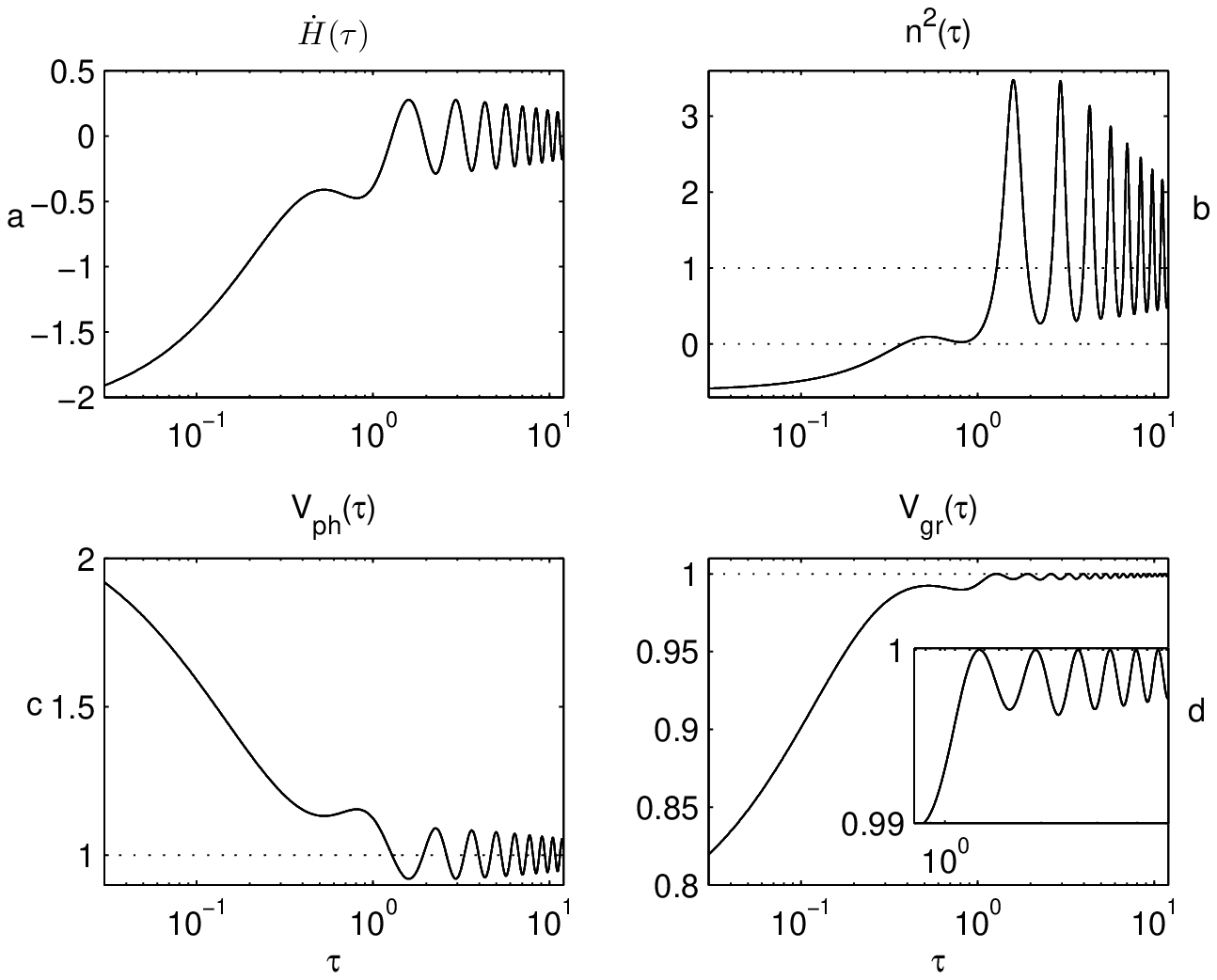}}
\caption {{\small Quasiperiodic universe.
The universe history is divided into a great
number of epochs by a finite number of transition points. Panel
(a) of $\dot{H}(t)$ shows clearly this behavior. The plot of
$n^2(t)$, see the panel (b) with $Q=2$, demonstrates that
in addition to an unlighted epoch of the first type,
arising at $t=0$,
an unlighted epoch of the second type
compressed into a point also appears.
This model shows also an unusual behavior of the phase
velocity in the early universe,
namely, it is greater than the speed of light
in vacuum during the first era of the first epoch, in contrast to the
models described above, and then tends to the asymptotic value equal
to one in the quasioscillatory regime. The behavior of the group
velocity is of a different kind, namely,
in the early universe it grows
monotonically, and then tends asymptotically to one in a
quasioscillatory regime, see panels (c) and (d) with $Q=0.3$.}}
\label{evolutionarymfig}
\end{figure}

\subsubsection{Three transition points universe}
\label{threetrans}
The fifth class of models is the class of universes with two epochs of
decelerated expansion and two epochs of accelerated evolution.  This
class arises when the acceleration parameter $-q(t)$ is of the
so-called ${\cal N}$-type \cite{Arch2}.
This model is also distinguished by the fact that the start of the
universe's expansion relates to the deceleration epoch, which is then
replaced by a short acceleration epoch. The second and final
accelerated epoch is of the de Sitter type.
We do not plot $-q(t)$, $H(t)$, $\rho(t)$, and
$\Pi(t)$, see \cite{Arch2} for such plots.
We are interested in light propagation and unlighted epochs.
Fig.~\ref{threetranspoin} illustrates this class of models.  The
unlighted epoch harbors now the first and second epochs of
deceleration, as well as the first epoch of accelerated
expansion. Clearly, this is a new feature.

\subsubsection{Quasiperiodic universe}
\label{evolutionarym}
The sixth class of models is the class of universes with a large
albeit finite number of transition points. It is a quasiperiodic
universe. This model is intermediate between the model with two
transition points and the periodic model.

This class arises when, starting from some transition point, the curve
${-}q(t)$ remains above the line $q{=}0$. This means that at a later
time the universe's expansion is accelerated. We do not plot $-q(t)$,
$H(t)$, $\rho(t)$, and $\Pi(t)$, see \cite{Arch2} for such plots.

We are interested in light propagation and unlighted epochs.
Fig.~\ref{evolutionarymfig} illustrates this class of models.
The behavior of the functions $\dot{H}(t)$,
$n^2(t)$, $V_{{\rm ph}}(t)$ and $V_{{\rm gr}}(t)$ is
quasiperiodic and the amplitudes of their oscillations decrease
asymptotically.  In addition to the unlighted epoch of the first type, one
can see now one point with $n^2=0$, i.e.,
a compressed unlighted epoch of the
second type. In between them the square of the refraction index is
positive. The behavior of the phase
velocity is unusual in the early universe since it is greater than the
speed of light in vacuum during the first era of the first epoch, and
then tends to the asymptotic value equal to one in a quasioscillatory
manner.
\begin{figure}
[t]
\centerline{\includegraphics[width=11.00cm,height=6.55cm]{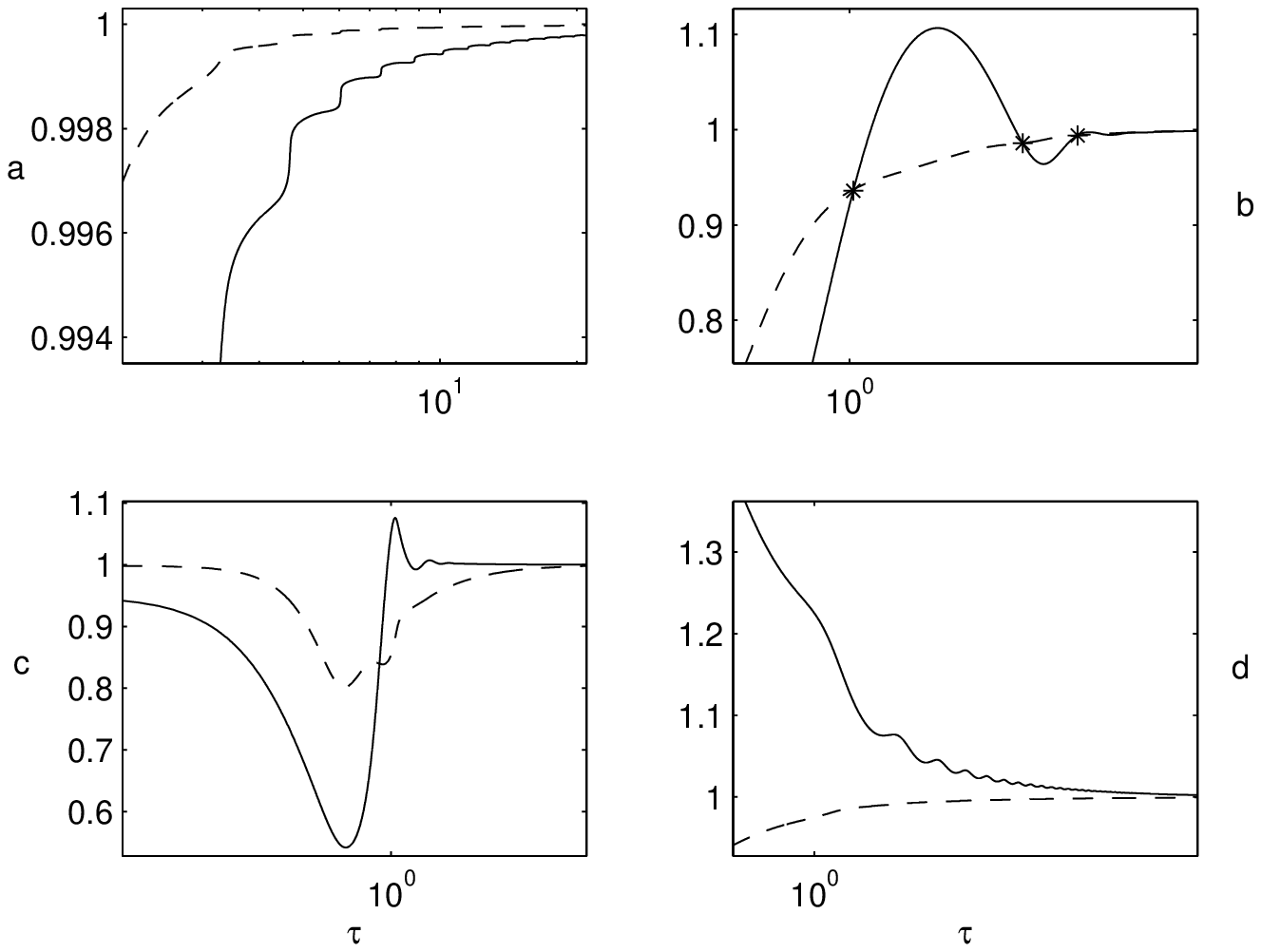}}
\caption {{\small Solid lines illustrate the behavior of
$\Gamma_{{\rm ph}}(t)$ and dashed lines illustrate the behavior of
$\Gamma_{{\rm gr}}(t)$.  $\Gamma_{{\rm ph}}(t)$ gives the ratio
between the phase length, which the electromagnetic wave runs during
a time $t$, if we take into account its phase velocity, and the
length traveled by a photon moving with the speed of light in pure
vacuum. For $\Gamma_{{\rm gr}}(t)$, the phase velocity is replaced
by the group velocity.  Panel (a) illustrates the periodic model, panel
(b) relates to the model with perpetual acceleration, panel (c)
contains the illustration to the model with one transition point, and
panel (d) illustrates the quasiperiodic model.  For the periodic and
quasiperiodic models the plots of the functions $\Gamma_{{\rm ph}}(t)$ and
$\Gamma_{{\rm gr}}(t)$ have no intersection points.
The other two models show the presence of crossing points. Generally,
in the late-time universe the quantities $\Gamma_{{\rm ph}}(t)$ and
$\Gamma_{{\rm gr}}(t)$ give practically the same results, the
ratios are
close to one. The difference between these quantities is,
however, essential
in the early universe.}}
\label{ratiofig}
\end{figure}

\subsubsection{Generic features of the
six models: The photon pathlength, the true duration
of epochs and the universe lifetime}
\label{photonpath}

Let us define two dimensionless functions,
\begin{equation}
\Gamma_{{\rm ph}}(t) \equiv \frac{1}{t} \ \int_0^t dt' \ V_{{\rm ph}}(t')
\,,
\label{L12}
\end{equation}
and
\begin{equation}
\Gamma_{{\rm gr}}(t) \equiv \frac{1}{t} \
\int_0^t dt' \ V_{{\rm gr}}(t') \,.
\label{L120}
\end{equation}
The first of them, $\Gamma_{{\rm ph}}(t)$, gives the ratio between
two quantities: the length traveled during a time $t$ by an
electromagnetic wave with phase velocity $ V_{{\rm ph}}(t)$
and the length traveled during the same time $t$ by a
photon moving with the speed of light in pure vacuum. In the second
quantity, $\Gamma_{{\rm gr}}(t)$, the phase velocity is replaced by
the group velocity.  Alternatively, one can consider these two
functions as mean values of the phase and group velocities averaged
over time $t$, respectively.
We have calculated these two functions for the models
discussed above. The results are presented in the Fig.~\ref{ratiofig}.
A general feature of all presented plots is that as $t \to \infty$
both these functions tend to one $\lim_{t \to \infty} \Gamma_{{\rm ph}}(t)
=1$ and $\lim_{t \to \infty} \Gamma_{{\rm gr}}(t) =1 $.
This is not surprising, since when $t \to \infty$ the contributions of
the periods with $V_{{\rm ph}} \neq 1$ and $V_{{\rm gr}}\neq 1$,
which in themselves are short, become vanishingly small. Thus, for
the estimation of the total life time of the universe we can assume
that, on average, photons propagate with speed of light in pure
vacuum. On the other hand, when we have to calculate the duration of
epochs and eras in the early universe, the behavior of $ \Gamma_{{\rm
ph}}$ and $ \Gamma_{{\rm gr}}$, as well as
the estimation of the photon pathlengths, depend on
the type and parameters of the model, and on whether one uses the
phase velocity or the group velocity.

\section{Conclusions}
\label{conc}

We can draw several conclusions.

1. The standard, concordant, cosmological model deals with at least
two epochs, during which electromagnetic waves cannot scan the
universe's internal structure neither bring information to
observers. The first epoch is when photons are in local thermodynamic
equilibrium with other particles, and the second is when photon
scattering by charged particles is strong. One can call these two
periods of cosmological time as standard unlighted epochs.  After the
last scattering surface, photons become relic photons and turn into a
source of information about the universe.

Now, if one takes into account electromagnetic interactions with the
dark sector, i.e., with the dark energy and dark matter, one can
expect that unlighted epochs of a new type can appear. Here we
described one possible example, namely, the unlighted epochs produced
by the non-minimal coupling of gravitational and electromagnetic
fields. Since in the framework of the non-minimal three-parameter
Einstein-Maxwell model, the curvature coupling can be formulated in
terms of an effective refraction index $n(t)$, as we did, we can take
advantage of the well-known classical analogy, namely, in a medium
with $n^2<0$ electromagnetic waves do not propagate and their group
velocity, i.e., energy transfer velocity, has zero value at the
boundary of the corresponding zone. Keeping in mind such an analogy we
studied here, both analytically and numerically, cosmological models
admitting unlighted epochs, the photon coupling to the spacetime
curvature being the key element of the models.

2. We established a formula for the group velocity, or energy transfer
velocity, of an electromagnetic wave non-minimally coupled to the
gravity field, namely, $V_{{\rm gr}}(t){=} \frac{2n(t)}{n^2(t){+}1}$,
with $|V_{{\rm gr}}(t)| \leq 1$.
Since unlighted epochs with a negative effective refraction index
squared can start and finish only when $n^2(t_{(*)}){=}0$ or
$n^2(t_{(**)}){=}\infty$, the group velocity is zero at the
unlighted epochs boundary points. At these points the electromagnetic
energy transfer stops, and thus the condition $V_{{\rm gr}}(t){=}0$
is the condition to be employed as the criterion for recognizing the
beginning and the end of the unlighted epochs.

We have distinguished three types of unlighted epochs: the first one
starts at $t{=}0$ with negative effective refraction index squared and
finishes when $n^2{=}0$; the second one starts at $t{=}t_{(1)}>0$ and
finishes at $t{=}t_{(2)}>t_{(1)}$ with $n^2(t_{(1)}){=}n^2(t_{(2)}){=}0$;
the third one is characterized by the fact that at least at one of the
two boundaries $n^2$ has an infinite value. The phase velocity
$V_{{\rm ph}}(t)$, another important physical characteristic of an
electromagnetic wave, is such that $V_{{\rm ph}}(t){=}\frac{1}{n(t)}$,
and so, when $n^2{=}0$ it can take infinite values at the beginning or
at the end of unlighted epochs.
The appearance of unlighted epochs of all three types were illustrated
analytically, using a cosmological model with a scale factor of the
Kohlrausch-type, i.e., a stretched exponential scale factor.

The unlighted epochs of the first and second types were then described
numerically using our Archimedean-type interaction model between dark
energy and dark matter. Clearly, the appearance or absence of
unlighted epochs is connected with the signs and absolute values of
the non-minimally coupling parameters $q_1$, $q_2$ and $q_3$. Thus,
fingerprints of the unlighted epochs in the cosmic microwave
background data could give some constraints on these non-minimal
coupling parameters. We hope to discuss this problem in another work.

3. The appearance of unlighted epochs caused by the non-minimal
coupling of photons to the gravitational field adds new features
into the history of the universe as written by electromagnetic fields.
Note that if the universe passes through an unlighted epoch of the
second type, we know for certain that information scanned by
electromagnetic waves during the preceding epochs is washed out, i.e.,
such unlighted epochs act as informational laundry.

In this connection there is a very interesting question: When and why
unlighted epochs of the second type can appear? Let us note, that
quite generally, the behavior of the function $n^2(t)$ inherits the
features of the function $\dot{H}(t)$; to see this compare the panels
(a) and (b) in Figs.~\ref{pepau}--\ref{evolutionarymfig}. In this
sense, the example of the model with two transition points, see
Fig.~\ref{twotranspoin}, shows explicitly that the appearance of
an unlighted epoch of the second type is a consequence of large
quasioscillations of the function $\dot{H}(t)$ and thus a consequence
of large quasioscillations of the spacetime curvature. Such
quasioscillations, in their turn, are the result of the dark matter
and dark energy energy-momentum redistribution, induced by the
Archimedean-type coupling \cite{Arch1,Arch2}.

4. The time span of different eras in the universe history can be
estimated using optical information.  From these durations one can
estimate the distance traveled by the electromagnetic waves through
their velocity of propagation.  But then, one should clarify what is
the electromagnetic wave propagation velocity which should be used in
the interpretation of the observational data.  Is it the group
velocity, the phase velocity, or the velocity of light in pure vacuum?
As we have seen these quantities differ in the early universe in an
essential way.  From Fig.~\ref{ratiofig} it is seen, for instance,
that the group velocity and the corresponding traveled distance are
less sensitive to the details of the dynamics of the universe.

\section*{Acknowledgments}

This work was partially funded by Funda\c c\~ao para a Ci\^encia e
Tecnologia (FCT) - Portugal, through Projects PTDC/FIS/098962/2008 and
PEst-OE/FIS/UI0099/2011.  AB is grateful to colleagues from CENTRA
(IST/UTL) for hospitality, and thanks the Russian Foundation for Basic
Research (Grants Nos. 11-02-01162-a and 11-05-97518-p-center-a), and
the FTP Scientific and Scientific-Pedagogical Personnel of the
Innovative Russia (Grants No. 16.740.11.0185 and 14.740.11.0407) for
support.

\section*{Appendix}
\label{Appendix}

As mentioned in  subsection \ref{refindex},
here we give several examples of specific non-minimal theories
by motivating possible
choices for the non-minimal coupling
parameters $q_1$,
$q_2$ and $q_3$.
Using the two effective non-minimal
coupling constants, $Q_1$ and $Q_2$ given by
\begin{equation}
Q_1 \equiv - 2(3q_1 {+} 2q_2 {+}
q_3) \,, \quad Q_2 \equiv -2 (3q_1 {+} q_2)  \,,
\label{R3a}
\end{equation}
we have shown we can write
the square of the refraction index in terms of Hubble
function $H(t)$ and acceleration parameter ${-}q(t)$ as
\begin{equation}
n^2(t) = \frac{1 + \left[Q_2 - Q_1 q(t)
\right]H^2(t) }{1 + \left[Q_1 - Q_2 q(t) \right] H^2(t)} \,.
\label{R49a}
\end{equation}
Of course $n^2(t)$ depends on $Q_1$ and $Q_2$, and thus on
$q_1$, $q_2$ and $q_3$.
However, independently of this choice,
for the de Sitter universe,
one obtains $n^2(t)=1$. Indeed,
since the acceleration parameter is constant
and is equal to one, $-q(t) {=} 1$, we obtain that $\varepsilon(t)=
\frac{1}{\mu(t)}= 1 {+} \left(Q_2 {+} Q_1 \right)H^2(t)$.
Although $\varepsilon(t)\neq 1$ and $\mu(t)
\neq 1$, one has $n^2(t) \equiv
\varepsilon(t){\mu(t)}= 1$ for arbitrary coupling parameters.

In general, the cosmological models are sensitive
to the signs and the values of the parameters $Q_1$ and $Q_2$.
There is an infinite variety
for the choice. However, six phenomenological
possibilities using geometric analogies and physical motivations
can be implemented.

\subsubsection*{Vanishing  non-minimal
susceptibility scalar (${\cal R}=0$)}
\label{ev}

When $6q_1{+}3q_2{+}q_3=0$ and thus, from Eq.~(\ref{susca}),
${\cal R}=0$, one has $Q_1+Q_2=0$. Putting $Q\equiv
Q_1=-Q_2$, one obtains in this one-parameter example
the following expression,
\begin{equation} n^2(t) = \frac{1 - Q H^2(t)\left[1+ q(t)\right]}{1 +
Q H^2(t)\left[1+ q(t)\right]} = \frac{1 + Q \dot{H}(t)}{1 - Q
\dot{H}(t)}\,.
\label{R43_x}
\end{equation}
This example has attracted our attention and we have used
it in our numerical calculations in Section \ref{num}.
Indeed, the interest in this
choice is that
only one function, $\dot{H}$,
guides the behavior of the effective refraction index, with $n^2 \to
1$, when $\dot{H} \to 0$. From a physical point of view this means
that asymptotically electromagnetic waves coupled non-minimally to
the gravitational field propagate with a phase velocity equal to
the speed
of light in vacuum.

\subsubsection*{Gauss-Bonnet-type relation}

When one imposes that the differential equations forming the
non-minimal Einstein-Maxwell system are of second order (see,
e.g., \cite{Horn,MHS}), then one should
use $q_1{+}q_2{+}q_3=0$ and
$2q_1{+}q_2=0$.
In this case the non-minimal susceptibility
tensor ${\cal R}_{ikmn}$ is proportional to the double dual Riemann
tensor $^{*}R^{*}_{ikmn}$, i.e., ${\cal R}_{ikmn} = \gamma \,
^{*}R^{*}_{ikmn}$ for some constant $\gamma$ \cite{bl05}.
Then we have $Q_1 = 0$, $Q_2= -2 q_1$,
and
\begin{equation}
n^2(t) = \frac{1 -2 q_1 H^2(t)}{1 + 2q_1 q(t) H^2(t)} \,.
\label{R5}
\end{equation}
If the non-minimal parameter $q_1$ is negative, then during the
acceleration epoch (i.e., $-q(t)>0$) the square of the refraction index is
positive, and there are no unlighted epochs.

\subsubsection*{First Weyl-type relation}

Assume now that ${\cal R}^{mn}=0$, in which case one can
take $3q_1+q_2=0$
and $q_2+q_3=0$.
Then, the susceptibility tensor is
proportional to the Weyl tensor ${\cal C}_{ikmn}$, i.e., ${\cal
R}_{ikmn} = \omega\,{\cal C}_{ikmn}$, and
$Q_1 = 0$, $Q_2= 0$. In this case
the effective permittivity scalars are equal to the unity,
$\varepsilon(t)=1$ and $\mu(t)=1$. Thus $n^2(t)=1$ in any
epoch.

\subsubsection*{Second Weyl-type relation}

Take $3q_1{+}q_2=0$ and $q_3=0$. Then
the susceptibility tensor is proportional to the
difference between the Riemann and Weyl tensors, i.e., ${\cal
R}_{ikmn} = \Omega [R_{ikmn} - {\cal C}_{ikmn}]$. In this example $Q_1
{=} 6 q_1$, $Q_2{=} 0$, and thus \begin{equation} n^2(t) = \frac{1 -6
q_1 q(t) H^2(t)}{1 + 6q_1  H^2(t)} \,.  \label{R541} \end{equation}
Now the square of the refraction index is positive during the
acceleration epoch ($-q(t)>0$) when the non-minimal parameter $q_1$ is
positive.

\subsubsection*{Symmetry relation with respect to the left and right
dualizations}

If one imposes that the left and right dual tensors coincide
${^{*}}{\cal R}^{ikmn}{=}{\cal R}^{*ikmn}$,
then one gets the example in which
$q_2{+}q_3=0$.  In this example $Q_1{=}Q_2$, and thus $n^2(t)=1$ for
arbitrary epochs, although the dielectric permittivity and magnetic
permeability \begin{equation} \varepsilon(t) = \frac{1}{\mu(t)} = 1 -
2(3q_1+q_2) H^2 [1-q(t)]  \label{R547} \end{equation} differ from
unity when $3q_1{+}q_2\neq 0$.

\subsubsection*{Drummond-Hathrell-type relation}

An example of a calculation for the three coupling parameters
based on one-loop corrections to quantum electrodynamics in curved
spacetime has been considered by Drummond and Hathrell \cite{Drummond}. In
this example the non-minimal coupling parameters are connected by the
relations $q_1\equiv-5 \tilde{Q}$, $q_2=13 \tilde{Q}$, $q_3=-2
\tilde{Q}$. The parameter $\tilde{Q}$ is positive, and constructed by
using
the fine structure constant $\alpha$ and the Compton wavelength of
the electron $\lambda_{\rm e}$, i.e., $\tilde{Q} \equiv
\frac{\alpha\lambda^2_{\rm e}}{180\pi}$. So, $Q_1= -18\tilde{Q}$,
$Q_2= 4\tilde{Q}$, yielding
\begin{equation}
n^2(t) = \frac{1 + 2 \tilde{Q} H^2(t)[2+9q(t)]}{1 - 2
\tilde{Q} H^2(t)[9-2q(t)]} \,.
\label{R51}
\end{equation}
This example was discussed in \cite{Tess}.

\end{document}